\documentclass[a4paper,11pt]{article}

\usepackage{jheppub} % for details on the use of the package, please
\usepackage{bbm}
\usepackage{dsfont}
\usepackage{youngtab}
\newcommand*{\longhookrightarrow}{\ensuremath{\lhook\joinrel\relbar\joinrel\rightarrow}}

\title{\boldmath Even spin $\mathcal{N}=4$ holography}
\author{Kevin Ferreira}
\affiliation{Institut f\"ur Theoretische Physik, ETH Z\"urich\\ CH-8093 Z\"urich, Switzerland}
\emailAdd{kferreira@itp.phys.ethz.ch}
\abstract{A two-dimensional Sp($2N$) vector model with small $\mathcal{N}=4$ superconformal symmetry is formulated, and its chiral algebra is shown to be freely generated by superprimary fields of even conformal weight. This vector model is the large level limit of a coset theory with large $\mathcal{N}=4$, whose proposed AdS$_3$ dual is a minimal Vasiliev higher spin theory with gauge algebra generated by fields of even spin. The relation of this vector model to the symmetric product orbifold, dual to tensionless strings in AdS$_3$ $\times$ S$^3$ $\times$ $\mathbb{T}^4$, is also worked out.}

\begin{document} 
\maketitle
\flushbottom
%%%%%%%%%%%%%%%%%%%%%%%%%%%%%%%%%%%%%%%%%%
\section{Introduction}
%%%%%%%%%%%%%%%%%%%%%%%%%%%%%%%%%%%%%%%%%%

Recent attemps in understanding the full scope of the AdS/CFT correspondence have put forward a relation between string theories and Vasiliev higher spin theories. The holographic duality establishes a correspondence between the free (or almost free) point of the conformal field theory and the high energy (tensionless) regime of string theory in AdS space. This regime is only partially understood \cite{Gross:1987ar,Gross:1987kza}, so further studies along these lines are expected to provide insights into the structure of string theory itself, see \cite{Sundborg:2000wp,Wittena,Sezgin:2002rt}.\\

In the high energy regime, the massive states of arbitrarily high spin which are present in string theory become massless, and it is believed that this signals the emergence of an unbroken phase of the theory, with enhanced symmetries generated by the massless higher spin states. Interacting theories of massless fields of arbitrarily high spin in AdS were constructed by Vasiliev, see e.g.\ \cite{Vasiliev:2003ev,Vasiliev:1999ba} and references therein, and it is therefore believed that string theory in AdS can be consistently restricted to a higher spin subsector described by the Vasiliev system of equations, in the tensionless regime. The precise way in which such a description arises in the context of string theory has not been completely elucidated yet. Nevertheless, holography gives important results, starting with a series of higher spin/CFT dualities \cite{Klebanov:2002ja} relating Vasiliev higher spin theories on AdS$_4$ to O($N$) vector models in three dimensions. Further work on such dualities include \cite{Giombi:2013fka,Sezgin:2003pt,Giombi:2010vg,Giombi:2009wh,Giombi:2012ms}, generalised to any number of dimensions in \cite{Sleight:2016xqq,Sleight:2016dba}, as well as the subsequent cases of \cite{Aharony:2011jz,Giombi:2011kc}, the 3$d$/2$d$ cases of \cite{Gaberdiel:2010pz,Ahn:2011pv,Gaberdiel:2011nt,Kumar:2016hpe,Gaberdiel:2012uj,Gaberdiel:2013vva,Gaberdiel:2014vca}, and the interesting dS$_6$/CFT$_5$ case of \cite{Fei:2015kta}.\\

Furthermore, an embedding of a higher spin theory into string theory was proposed in \cite{Chang:2012kt}, in which the $\mathcal{N} = 6$ U($N$)$\times$U($M$) ABJ theory is related to a higher spin theory with U($M$) Chan-Paton indices. Since the ABJ theory is believed to be dual to a string theory in AdS$_4$, this proposal establishes a triality between a higher spin theory, a string theory, and the dual CFT. The higher spin/CFT dualities appear when $M$ is finite and $N$ is large, at a point where the bulk coupling $\lambda_{bulk}\sim M/N$ is small, whereas the strings/CFT duality arises in the regime $M\sim N$ large. The intuitive picture elaborated in \cite{Chang:2012kt} is that strings correspond to the flux tubes of the U($M$) higher spin theory, which appear at strong coupling.\\

The 3$d$/2$d$ case is considerably different and has been proposed in \cite{Gaberdiel:2014cha,Gaberdiel:2015mra,Gaberdiel:2015wpo,Gaberdiel:2017ede}, building on the 3$d$ higher spin/CFT$_2$ duality of \cite{Gaberdiel:2013vva}. In this case, the 't Hooft limit of a 1-parameter family of $\mathcal{N}=4$ coset models \cite{Sevrin:1988ew,Schoutens:1988ig,Spindel:1988sr,VanProeyen:1989me} is related to a 1-parameter family of higher spin theories in AdS$_3$ with $2\times 2$ Chan-Paton factors, with parameter $\lambda$. At $\lambda =0$, the coset becomes a free-field vector model, which can be embedded in the symmetric product theory
\begin{equation}\label{eqn:symm_prod}
\left(\mathbb{T}^4\right)^{N+1}/S_{N+1} \; ,
\end{equation}
believed to be dual to string theory on AdS$_3$ $\times$ S$^3$ $\times$ $\mathbb{T}^4$ at the tensionless point \cite{David:2002wn,Gaberdiel:2014cha}. This embedding is a concrete manifestation of the idea that string theory in the tensionless regime has a higher spin subsector. Indeed, the untwisted sector of the symmetric product orbifold can be completely decomposed in terms of a single representation, and all its tensor powers, of an emergent symmetry algebra, the so-called higher spin square \cite{Gaberdiel:2015mra}, which is generated by two independent higher spin symmetries.\\

In addition to the $\mathcal{N}=4$ cosets studied in \cite{Gaberdiel:2013vva}, there are other coset models with the same supersymmetry rank, as listed in \cite{Sevrin:1989ce}.
These could then be used to perform a construction similar to the one of \cite{Gaberdiel:2014cha,Gaberdiel:2015mra,Gaberdiel:2015wpo,Gaberdiel:2017ede}, and in this way find new relations between string theory and higher spin theories.
Only two of the cosets in \cite{Sevrin:1989ce} are promptly susceptible to be dual to a standard Vasiliev higher spin theory: the SU-type coset of \cite{Gaberdiel:2013vva}, and a Sp-type coset, expressed below in equation \eqref{eqn:COSET}.
There are a few arguments sustaining this view.
First, as usual in holographic dualities, the CFT$_2$ is expected to admit a large $N$ expansion.
All $\mathcal{N}=4$ cosets which do not have such a parameter are therefore discarded in a first analysis.
We are left with the SU-type and Sp-type cosets mentioned above, as well as an SO-type coset.
Nevertheless, this SO-type coset has an unwanted property: due to an $\mathfrak{su}(2)$ factor in the denominator, its chiral algebra is not freely generated even in the 't Hooft limit.
This makes this coset unsuited to a comparison with the classical standard Vasiliev higher spin theories, which are described by a freely generated algebra.
It is possible that this SO-type coset is dual to a minimal Vasiliev theory with modified boundary conditions, in the spirit of \cite{Candu:2013fta}.
This issue will be studied elsewhere.\\

With these considerations, it is natural to study the $\mathcal{N}=4$ coset described by
\begin{equation}\label{eqn:COSET}
\frac{\mathfrak{sp}(2N+2)_{k+N+2}^{(1)}}{\mathfrak{sp}(2N)_{k+N+2}^{(1)}}\oplus\mathfrak{u}(1)^{(1)} \; .
\end{equation}
In this paper we establish and study the holographic duality between a minimal Vasiliev higher spin theory in AdS$_3$ with $2\times 2$ Chan-Paton factors and the coset model \eqref{eqn:COSET} in the 't Hooft limit.
The chiral algebra of the coset in the 't Hooft limit, and correspondingly the gauge sector of the dual higher spin theory, is spanned by one superprimary field per even conformal weight, thus extending the previous bosonic even spin construction \cite{Ahn:2011pv,Gaberdiel:2011nt,Candu:2012ne}, the $\mathcal{N}=2$ case \cite{Ferreira:2014xsr}, and a recent $\mathcal{N}=1$ proposal \cite{Kumar:2016hpe}.\\ 

In the $k\rightarrow\infty$ limit, the coset \eqref{eqn:COSET} is described by an Sp(2$N$) vector model. The results of \cite{Gaberdiel:2014cha} can then be used to obtain the relation between the Sp($2N$) vector model and the symmetric product theory \eqref{eqn:symm_prod}, thus providing another possible description of the embedding of higher spins in string theory. In particular, the higher spin symmetry we find in this case can be embedded into that of \cite{Gaberdiel:2014cha}, and this allows us to construct the stringy symmetry algebra using similar arguments. In this way, the untwisted sector of the symmetric product orbifold can be decomposed further into representations of the untwisted sector of the Sp$(2N$) vector model.\\

It is interesting to notice that, in contrast with previous even spin constructions \cite{Klebanov:2002ja,Ahn:2011pv,Gaberdiel:2011nt,Ferreira:2014xsr,Candu:2012ne}, the $\mathcal{N}=4$ version is obtained using a Sp(2$N$) rather than a O($N$) model. In the $2d$ bosonic case \cite{Ahn:2011pv,Gaberdiel:2011nt}, the bosonic even spin $\mathcal{W}$-algebra was related to different orthogonal and symplectic models \cite{Candu:2012ne}, which are described at finite coupling by various cosets whose chiral algebra is freely generated.
Even though an analoguous analysis of the quantum $\mathcal{N}=4$ even spin $\mathcal{W}$-algebra is beyond the scope of this paper, we will briefly comment on this issue.\\

This paper is organised as follows. In Section \ref{sec:EVEN} we construct a family of two-dimensional theories with $\mathcal{N}=4$ superconformal symmetry. We start with an Sp($2N$) vector model and proceed to find its chiral algebra. Then we turn on a coupling $\lambda$ which introduces interactions between the fields. For general $\lambda$, the theory is described by the 't Hooft limit of the coset model \eqref{eqn:COSET}. In section \ref{sec:HS} we construct the gauge algebra of the higher spin Vasiliev theory on AdS$_3$, proposed as holographic dual to the coset CFT. This is achieved via a truncation of an extended higher spin algebra, whose spectrum is found and seen to match with the chiral spectrum of the CFT. The massive spectrum is also computed, and used to match one-loop partition functions in Appendix \ref{app:DUAL}. In section \ref{sec:SYMM} we elaborate on the relation between the Sp(2$N$) vector model and string theory, building on the results of \cite{Gaberdiel:2014cha} on the symmetric product orbifold. Finally, various conventions and technical details are collected in the appendices.

%%%%%%%%%%%%%%%%%%%%%%%%%%%%%%%%%%%%%%%%%%
\section{Even spin $\mathcal{N}=4$ $\mathcal{W}$-algebra}\label{sec:EVEN}
%%%%%%%%%%%%%%%%%%%%%%%%%%%%%%%%%%%%%%%%%%

In this section we present a two-dimensional CFT whose chiral algebra realises a $\mathcal{W}_{\infty}$ algebra with generators of even conformal weight, and with the $\mathcal{N}=4$ superconformal algebra as a subalgebra (c.f.\ \cite{Gaberdiel:2013vva} for a revision of the $\mathcal{N}=4$ superconformal algebra). We start with a simple setting consisting of a symplectic vector model of free bosons and free fermions at large $N$, with so-called small $\mathcal{N}=4$ symmetry, and organise the spectrum of generators according to representations of the superconformal algebra. At finite coupling, the vector model is described by a coset CFT which has the so-called large $\mathcal{N}=4$ symmetry, and possesses the same set of chiral generators.
%%%%%%%%%%%%%%%%%%%%%%%%%%%%%%%%%%%%%%%%%%
\subsection{The Sp($2N$) vector model}
%%%%%%%%%%%%%%%%%%%%%%%%%%%%%%%%%%%%%%%%%%

The Sp($2N$) vector model with small $\mathcal{N}=4$ consists of $4N$ fermionic and bosonic fields transforming as
\begin{equation}\label{eqn:REPSR}
\begin{aligned}
\text{bosons: } &  {\bf 2N}_{({\bf 1},{\bf 2})} \\
\text{fermions: } & {\bf 2N}_{({\bf 2},{\bf 1})} \; ,
\end{aligned}
\end{equation}
where ${\bf 2N}$ denotes the vector representation of Sp($2N$), and the subscripts label the quantum numbers with respect to two global symmetry algebras $(\mathfrak{su}(2)_+,\mathfrak{su}(2)_-)$. These global algebras constitute the R-symmetry of large $\mathcal{N}=4$.\footnote{To make contact with the coset model, which has large $\mathcal{N}=4$, we will keep track of the quantum numbers of both $\mathfrak{su}(2)_{\pm}$. In reality, the R-symmetry of the small $\mathcal{N}=4$ superconformal algebra is given by $\mathfrak{su}(2)_+$ alone.} We denote the NS fermions as $\psi^{i,\alpha}$, and the bosons as $\mathcal{J}^{i,\beta}$. Here $i=\pm 1,\ldots, \pm N$ is a vector index (c.f.\ appendix \ref{app:coset} for conventions), $\alpha=\pm$ labels the ${\bf 2}$ of $\mathfrak{su}(2)_+$, and  $\beta =\pm$ the ${\bf 2}$ of $\mathfrak{su}(2)_-$. The conserved currents of the vector model are given by the Sp($2N$) invariant combinations of these fields (see appendix \ref{subsec:VAC} for a more detailed derivation of the spectrum). Using just the bosons, conserved currents of conformal weight $s$ are constructed as
\begin{equation}
\Omega_{ij}\;\partial^r\mathcal{J}^{i,\beta_1}\; \partial^{s-2-r}\mathcal{J}^{j,\beta_2} \; ,
\end{equation}
where $\Omega$ is the 2$N$ $\times$ $2N$ symplectic matrix, and $r=0,\ldots,s-2$. Due to the anti-symmetry of $\Omega$, for odd $s\geq 3$ primary fields of this kind transform in the symmetric tensor product of $({\bf 1},{\bf 2})$ with itself. The anti-symmetric part is a descendant, as can be seen for example for $s=3$, since
\begin{align}
& \; \Omega_{ij}\left[ \mathcal{J}^{i,+}\; \partial \mathcal{J}^{j,-} - \mathcal{J}^{i,-}\; \partial \mathcal{J}^{j,+}\right] \nonumber \\ 
= & \; \Omega_{ij}\left[ \mathcal{J}^{i,+}\; \partial \mathcal{J}^{j,-} - \left(\partial \mathcal{J}^{j,-}\right) \mathcal{J}^{i,+} + \partial \left ( \mathcal{J}^{j,-} \mathcal{J}^{i,+} \right) \right]  \\ 
= & \;  \Omega_{ij}\partial \left ( \mathcal{J}^{j,-} \mathcal{J}^{i,+} \right) \; .\nonumber
\end{align}
Therefore for each odd $s\geq 3$ the primary fields transform as $({\bf 1},{\bf 3})$. On the other hand, for even $s\geq 2$ we pick the anti-symmetric self-product, which gives one primary field $({\bf 1},{\bf 1})$.\\

Using now the fermions consider, for $r=0,\ldots,s-1$,
\begin{equation}
\Omega_{ij}\; \partial^r\psi^{i,\alpha_1}\; \partial^{s-1-r} \psi^{j,\alpha_2} \; .
\end{equation}
In the same way as before, the resulting primary currents transform as $({\bf 3},{\bf 1})$ for each odd $s\geq 1$, and as $({\bf 1},{\bf 1})$ for even $s\geq 2$. Finally, the currents
\begin{equation}
\Omega_{ij}\; \partial^r\mathcal{J}^{i,\beta}\; \partial^{s-3/2-r} \psi^{j,\alpha} \; ,
\end{equation}
contribute with four primaries for each half-integer $s\geq 3/2$, transforming as $({\bf 2},{\bf 2})$. In total the chiral spectrum is then generated by
\begin{alignat}{3}
\begin{aligned}
s=1: & \qquad & ({\bf 3}, \, & {\bf 1})\\
s \text{ odd }: & \qquad & ({\bf 3},{\bf 1})\; \oplus &\; ({\bf 1},{\bf 3}) \\
s \text{ even }: & \qquad & ({\bf 1},{\bf 1})\; \oplus &\; ({\bf 1},{\bf 1}) \\
s \text{ half-integer}: & \qquad &  ({\bf 2}, \, & {\bf 2}) \;  ,
\end{aligned}
\end{alignat}
which can be organised in $\mathcal{N}=4$ multiplets as
\begin{equation}\label{eqn:CHIRAL}
(\mathcal{N}=4)\oplus\bigoplus_{n=1}^{\infty} R^{(2n)} \; ,
\end{equation}
where $(\mathcal{N}=4)$ stands for the small $\mathcal{N}=4$ superconformal algebra, generated by three $s=1$ currents, four supercharges at $s=3/2$, and the energy momentum tensor. The zero modes of the $s=1$ currents generate $\mathfrak{su}(2)_+$. Also, $R^{(s)}$ is the chiral $\mathcal{N}=4$ multiplet with lowest spin $s$, and with R-symmetry quantum numbers
\begin{alignat}{3}
\qquad & \qquad   & s: & \qquad & ({\bf 1}, &{\bf 1}) \nonumber\\
\qquad & \qquad   & s+1/2: & \qquad & ({\bf 2}, & {\bf 2}) \nonumber\\
R^{(s)}: & \qquad & s+1: & \qquad & ({\bf 3},{\bf 1}) \; \oplus &\; ({\bf 1},{\bf 3}) \\
\qquad & \qquad & s+3/2: & \qquad & ({\bf 2}, &{\bf 2}) \nonumber\\
\qquad & \qquad & s+2: & \qquad & ({\bf 1}, & {\bf 1} ) \nonumber \; .
\end{alignat}

This vector model then realises a $\mathcal{W}$-algebra with small $\mathcal{N}=4$ symmetry and $c=6N$, whose chiral spectrum contains only even spin superprimaries.\\

In order to make contact with the large $\mathcal{N}=4$ coset model of the next section, and the subsequent results, we add to this chiral algebra four free bosonic and four free fermionic fields transforming as
\begin{equation}\label{eqn:REPSR2}
\begin{aligned}
\text{bosons: } &  {\bf 1}_{({\bf 1},{\bf 3})}\oplus {\bf 1}_{({\bf 1},{\bf 1})}\\
\text{fermions: } & {\bf 1}_{({\bf 2},{\bf 2})} \; ,
\end{aligned}
\end{equation}
under Sp$(2N)$ and $(\mathfrak{su}(2)_+,\mathfrak{su}(2)_-)$. Upon turning on a level, the global $\mathfrak{su}(2)_-$ is generated by the zero modes of the bosonic currents. These additional fields correspond to the free currents obtained when contracting the large $\mathcal{N}=4$ to the small $\mathcal{N}=4$ algebra, c.f. \cite{Gaberdiel:2013vva}.\\

We will call this algebra $\mathcal{W}^{\text{e},\; \mathcal{N}=4}_{\infty}[0]$. It corresponds to an even-spin version of the 1-parameter family of large $\mathcal{N}=4$ $\mathcal{W}$-algebras denoted $\mathcal{W}^{\mathcal{N}=4}_{\infty}[\lambda]$, constructed in \cite{Beccaria:2014jra}, at $\lambda=0$. For generic $\lambda$ the structure constants of this algebra were shown in \cite{Beccaria:2014jra} to be completely fixed by two parameters $k^{\pm}$, corresponding to the levels of the affine $\mathfrak{su}(2)_{\pm}$ subalgebras of the large $\mathcal{N}=4$ superconformal algebra. These can be exchanged by the central charge $c$ and the parameter $\lambda$ as
\begin{alignat}{3}
c=\frac{6k^+k^-}{k^++k^-} \; ,& \qquad & \lambda = \frac{k^+}{k^++k^-} \; .
\end{alignat}
Note that $k^-\rightarrow\infty$ corresponds to $\lambda =0$, and $c=6k^+=6(N+1)$, corresponding to the vector model. In the same way, we expect $\mathcal{W}^{\text{e},\;\mathcal{N}=4}_{\infty}[0]$ to be the $\lambda=0$ point of a 1-parameter family of algebras $\mathcal{W}^{\text{e},\; \mathcal{N}=4}_{\infty}[\lambda]$, which are also completely determined by two levels $k^{\pm}$ for any value of $\lambda$. We do not construct this algebra explicitly here, but believe that there are good indications that this expectation is coherent. The foremost indication of this is the explicit coset realisation of $\mathcal{W}^{\text{e},\; \mathcal{N}=4}_{\infty}[\lambda]$ for positive integer values of $k^{\pm}$, namely $k^+=N+1$, $k^-=k+1$, constructed in the next section.

%%%%%%%%%%%%%%%%%%%%%%%%%%%%%%%%%%%%%%%%%%%%%%%%%%
\subsection{Coset generalisation}\label{subsec:COS}
%%%%%%%%%%%%%%%%%%%%%%%%%%%%%%%%%%%%%%%%%%%%%%%%%%

The Sp($2N$) vector model arises as the $k\rightarrow\infty$ limit of the coset theory given by
\begin{equation}\label{eqn:coset}
\frac{\mathfrak{sp}(2N+2)_k}{\mathfrak{sp}(2N)_{k+1}}\oplus\mathfrak{so}(4N+4)_1\oplus\mathfrak{u}(1) \; ,
\end{equation}
which was shown to have $\mathcal{N}=4$ superconformal symmetry in \cite{Sevrin:1989ce}. The $\mathfrak{so}(4N+4)_1$ factor encodes $4N+4$ fermions, which are free for any value of $k$. All the details of the construction of this coset can be found in Appendix \ref{app:coset}.\\

The representation theory of the coset is completely determined by the representation theory of each of its Ka$\check{\text{c}}$-Moody components. We will disregard the $\mathfrak{u}(1)$ factor by putting its momentum to zero. Furthermore, by construction, the NS free fermions are either in the vector or in the vacuum representation of $\mathfrak{so}(4N+4)_1$. Therefore, coset representations are labelled by a pair of representations $(\Lambda_+;\Lambda_-)$, where $\Lambda_+$ is a representation of $\mathfrak{sp}(2N+2)_k$, and $\Lambda_-$ is a representation of $\mathfrak{sp}(2N)_{k+1}$. \\

The central charge of the CFT defined by the coset is
\begin{equation}
c=\frac{6k^+k^-}{k^++k^-} \; ,
\end{equation}
where $k^+=N+1$ and $k^-=k+1$. A precise correspondence between the $k\rightarrow\infty$ limit of the coset and the vector model is found in Appendix \ref{app:kINFTY}, which builds on similar results in \cite{Gaberdiel:2014vca,Gaberdiel:2013vva,Gaberdiel:2011aa,Ferreira:2014xsr}. In particular, the untwisted sector of the vector model is captured by the $k\rightarrow\infty$ limit of the $(\Lambda_+;\Lambda_-)=(0;\Lambda)$ subsector of the coset representations, where $0$ denotes the trivial representation, and $\Lambda$ denotes a general representation of $\mathfrak{sp}(2N)$. Combining left- and right-movers, the Hilbert space of the untwisted sector is then
\begin{equation}\label{eq:UNTWIST}
\mathcal{H}_U=\bigoplus_{\Lambda}(0;\Lambda)\otimes\overline{(0;\Lambda^*)} \; ,
\end{equation}
where $\Lambda^*$ denotes the conjugate of $\Lambda$, and since representations of $\mathfrak{sp}(2N)$ are self-conjugate we have $\Lambda^*=\Lambda$. The sum runs over all representations $\Lambda$ which are obtained by taking successive tensor products of the vector representation, thus covering all representations of $\mathfrak{sp}(2N)$, see \cite{FultonHarris1991}. Denoting the vector representation as $v\equiv{\bf 2N}$, the minimal non-trivial representation of the untwisted sector is  $(0;v)$, and its conformal dimension is
\begin{equation}
h(0;v)=\frac{k+3/2}{2(k+N+2)}\xrightarrow{k\rightarrow\infty}\frac{1}{2} \; .
\end{equation}
It therefore corresponds to the state
\begin{equation}
\psi_{-1/2}^{i,\alpha}\vert 0\rangle \; ,
\end{equation}
of the vector model. The bosons of the vector model correspond to the superconformal descendants of $(0;v)$, see section \ref{subsec:MASSIVE}, and other similar cases in \cite{Gaberdiel:2013vva,Gaberdiel:2014cha}.\\

The 't Hooft limit of the coset theory is defined as $N,k\rightarrow\infty$ with
\begin{equation}
\lambda = \frac{k^+}{k^++k^-}= \frac{N+1}{k+N+2}\simeq \frac{N}{k+N}
\end{equation}
kept fixed. Note that the central charge can be expressed as
\begin{equation}
c=6\lambda k^-\; ,
\end{equation}
which diverges in this limit, unless $\lambda =0$, in which case we recover the vector model with $c=6(N+1)$. In the 't Hooft limit the chiral algebra of the coset CFT is freely generated, see Appendix \ref{app:DUAL}, and \cite{Candu:2013fta} for a similar discussion. The chiral fields of the coset theory are given by the chiral fields of the Sp($2N$) vector model corrected with terms proportional to $\lambda$. These terms ensure that their OPE's with the denominator currents are non-singular, c.f.\ \cite{Candu:2013fta} for a more detailed discussion about this point. These corrections do not change the counting of the fields, and therefore the chiral spectrum of the coset in the 't Hooft limit is also given by \eqref{eqn:CHIRAL}, with $(\mathcal{N}=4)$ now denoting the large $\mathcal{N}=4$ superconformal algebra.

%%%%%%%%%%%%%%%%%%%%%%%%%%%%%%%%%%%%%%%%%%%%%
\section{Higher spin dual}\label{sec:HS}
%%%%%%%%%%%%%%%%%%%%%%%%%%%%%%%%%%%%%%%%%%%%%

The AdS$_3$ gravitational theory dual to the coset model in the 't Hooft limit, and in particular to the Sp(2$N$) vector model, is constructed from the extended supersymmetric Vasiliev higher spin theory based on the gauge algebra $\mathfrak{shs}_2[\mu]$ by a consistent even-spin truncation. This truncation is performed using an involutive graded automorphism of $\mathfrak{shs}_2[\mu]$ (see \cite{Vasiliev:1986qx} for a  revision of these concepts, and Appendix \ref{app:HS} for a brief introduction to the necessary machinery). The dual to the vector model is obtained at $\mu=0$. The construction of this truncated higher spin theory and the necessary checks for its consistency as a dynamical system were obtained in \cite{Vasiliev:1999ba,Konstein:1989ij,Prokushkin:1998bq}, where its gauge algebra is denoted $husp(2,2\vert 4)$.\\

%%%%%%%%%%%%%%%%%%%%%%%%%%%%%%%%%%%%%%%%%%%%%%%%%%%
\subsection{Truncation of $\mathfrak{shs}_2[\mu]$}\label{subsec:TRUNC}
%%%%%%%%%%%%%%%%%%%%%%%%%%%%%%%%%%%%%%%%%%%%%%%%%%%

We are interested in automorphisms of the super Lie algebra $\mathfrak{shs}_2\left[\mu\right]$ which can be obtained by negating an anti-automorphism of the associative algebra $sB_2[\mu]=sB[\mu]\otimes$Mat($2,\mathbb{C}$), see Appendix \ref{app:HS}. An anti-automorphism of $sB_2[\mu]$ is obtained by  composing an anti-automorphism of $sB[\mu]$, denoted $\eta$, with an anti-automorphism of Mat($2,\mathbb{C})$, denoted $\rho$. The map $\rho$ defined as 
\begin{equation}
 \rho(M)= \Omega^{-1}M^t\Omega \; ,
\end{equation}
for $M\in$ Mat($2,\mathbb{C})$, where $\Omega$ is the $2\times 2$ symplectic matrix obeying $\Omega^{-1}=\Omega^t=-\Omega$, is an involutive anti-automorphism of Mat($2,\mathbb{C})$, since
\begin{alignat}{3}
\rho^2(M) = M \; ,& \qquad & \rho(\left[ M_1,M_2\right]) = -\left[\rho(M_1),\rho(M_2)\right] \; .
\end{alignat}
On the other hand, the action of $\eta$ on $sB[\mu]$ is defined via its action on the oscillators which realise the algebra, see Appendix \ref{app:HS}, which will be taken as
\begin{alignat}{3}
\eta(\hat{y}_{\alpha})=-\hat{y}_{\alpha}, & \qquad & \eta(k)=k, & \qquad & \eta(\mathds{1})=\mathds{1} \; ,
\end{alignat}
and which is involutive, and compatible with the defining relations \eqref{eqn:RELATIONS}. Combining $\eta$ with $\rho$ produces an involutive anti-automorphism of $sB_2[\mu]$, which preserves the $\mathbb{Z}_2$-grading, and an automorphism of $\mathfrak{shs}_2[\mu]$ can be constructed by negating it. We absorb the negation into $\tau\equiv-\eta$, so that the final automorphism $\tau_2$ of $\mathfrak{shs}_2[\mu]$ is
\begin{equation}
\tau_2\left( a\otimes M\right) = \tau(a)\otimes \Omega^{-1}M^t\Omega \; ,
\end{equation}
for $a\in\mathfrak{shs}[\mu]$, $M\in$ Mat($2,\mathbb{C})$, where $\tau$ is the negation of $\eta$,
\begin{alignat}{3}
 \tau(\hat{y}_{\alpha})=\hat{y}_{\alpha}, & \qquad & \tau(k)=-k, & \qquad & \tau(\mathds{1})=-\mathds{1} \; .
\end{alignat}

We can now construct the subalgebra $\mathfrak{shs}^{\mathfrak{sp}}_2[\mu]\subset \mathfrak{shs}_2[\mu]$, defined as the truncation which only keeps the elements $A\in\mathfrak{shs}_2[\mu]$ such that
\begin{equation}
\tau_2(A)=A \; .
\end{equation}

This truncation preserves the $D(2,1;\alpha)$ subalgebra of $\mathfrak{shs}_2[\mu]$, given in \eqref{eqn:D210}, with $\alpha = \mu/(1-\mu)$. To see this, notice that the action of $\tau$ on $sB[\mu]$ satisfies, for any $a_1,a_2\in sB[\mu]$,
\begin{align}\label{eqn:tau}
\begin{aligned}
\tau(a_1a_2) & = -\eta(a_1a_2) = -(-1)^{\vert a_1\vert\vert a_2\vert}\eta(a_2)\eta(a_1) \\
& = -(-1)^{\vert a_1\vert\vert a_2\vert}\tau(a_2)\tau(a_1) \; ,
\end{aligned}
\end{align}
where we used that $\eta$ is an anti-automorphism of $sB[\mu]$. From this we can deduce $\tau(\hat{y}_{\alpha}\hat{y}_{\beta})=\hat{y}_{\alpha}\hat{y}_{\beta}$, so that the $\mathfrak{sl}(2)$ generators $L_0$, $L_{\pm 1}$ are automatically preserved by $\tau_2$. Concerning $A_0^{\pm,i}$, we have
\begin{equation}
\tau_2(A_0^{\pm,i}) = -\frac{1}{2}(1\pm k)\otimes \Omega^{-1}(\sigma^i)^t\Omega \; ,
\end{equation}
and since $\mathfrak{sp}(2)$ preserves $\Omega$, i.e. $\sigma^i\Omega+\Omega(\sigma^i)^t =0$, we deduce 
\begin{equation}
\Omega^{-1}(\sigma^i)^t\Omega=-\sigma^i \; ,
\end{equation}
and finally $\tau_2(A_0^{\pm,i})=A_0^{\pm,i}$. Lastly, concerning the fermionic generators of $D(2,1;\alpha)$, we first find
\begin{align}
\begin{aligned}
\tau\left(\hat{y}_\alpha k\right) & = -\tau(k)\tau(\hat{y}_{\alpha}) = k\hat{y}_{\alpha}\\
& = -\hat{y}_{\alpha} k \;,
\end{aligned}
\end{align}
where we have used that $\hat{y}_\alpha k=-k\hat{y}_\alpha$. Together with the fact that $E_{12}$, $E_{21}$, and $(E_{11}-E_{22})$ preserve $\mathfrak{sp}(2)$, this leads to $\tau_2(G_r^{ab})=G^{ab}_r$. Therefore, $D(2,1;\alpha)\subset\mathfrak{shs}^{\mathfrak{sp}}_2[\mu]$.\\

%%%%%%%%%%%%%%%%%%%%%%%%%%%%%%%%%%%%%%
\subsection{Massless spectrum}
%%%%%%%%%%%%%%%%%%%%%%%%%%%%%%%%%%%%%%

The gauge sector of $\mathfrak{shs}_2[\mu]$ can be organised in representations of the subalgebra $D(2,1;\alpha)$ \cite{Gaberdiel:2013vva,Candu:2013fta,Candu:2014yva}. The states of highest spin in a $D(2,1;\alpha)$ multiplet are proportional to
 \begin{equation}
  \hat{y}_1^{2r+2}\otimes\mathbbm{1}_2 \; , 
 \end{equation}
for $r\in\mathbb{N}^0$, with $r=0$ giving $L_1$ itself. Note that $r$ is related to the $\mathfrak{sl}(2)$ spin $s$ as $s=r+1$, since $\hat{y}_1$ carries helicity 1/2. The highest weights surviving the truncation will be those for which $\tau(\hat{y}_1^{2r+2})=\hat{y}_1^{2r+2}$. For general $r$, using \eqref{eqn:tau},
\begin{align}
\tau(\hat{y}_1^{2r+2}) & = \tau(\hat{y}_1^{2r}\hat{y}_1^2) \nonumber \\
& = - \tau(\hat{y}_1^2)\tau(\hat{y}_1^{2r}) \\
& = - \hat{y}_1^2\tau(\hat{y}_1^{2r}) \nonumber \; .
\end{align}
We can now take one more step to find $\tau(\hat{y}_1^{2r})$ in the same way:
\begin{align}
 \tau(\hat{y}_1^{2r}) & = \tau(\hat{y}_1^{2r-2}\hat{y}_1^{2})=-\hat{y}_1^2\tau(\hat{y}_1^{2r-2})\;,
\end{align}
so that
\begin{equation}
\tau(\hat{y}_1^{2r+2}) = \hat{y}_1^4 \tau(\hat{y}_1^{2r-2})\; ,
\end{equation}
and so on until we reach $\tau(\mathds{1})$, and recall that $\tau(\mathds{1})=-\mathds{1}$. From this we deduce
\begin{equation}
\tau(\hat{y}_1^{2r+2}) =(-1)^r\hat{y}_1^{2r+2} \; .
\end{equation}

Therefore the only highest weight states of $D(2,1;\alpha)$ that generate $\mathfrak{shs}_2[\mu]$ and survive the truncation by $\tau_2$ are
\begin{alignat}{3}
  \hat{y}_1^{2r+2}\otimes\mathbbm{1}_2 \;, & \qquad & r\in 2\mathbb{N}^0 \; .
\end{alignat}
This corresponds to odd $\mathfrak{sl}(2)$ spin $s$.\footnote{By the usual AdS$_3$/CFT$_2$ relations, in the dual CFT this corresponds to even conformal weight.} Since we have explicitly shown that $D(2,1;\alpha)$ survives the truncation, then the whole multiplet generated from one surviving highest weight state also survives the truncation. In the same way, the whole multiplet generated by a highest weight of $D(2,1;\alpha)$ which does not survive the truncation has the same fate. This conclusion can be explicitly confirmed by a direct computation of the action of $\tau_2$ on the higher spin fields.\\

Recalling that $R^{(n)}$ denotes the $D(2,1;\alpha)$ multiplet with \emph{lowest} helicity $n=s-1$  (from a CFT$_2$ perspective), then
\begin{equation}
\mathfrak{shs}^{\text{sp}}_2[\mu] = D(2,1;\alpha)\oplus\bigoplus_{n=1}^{\infty} R^{(2n)} \; .
\end{equation}
This precisely matches the spectrum of superprimaries \eqref{eqn:CHIRAL} we found in the chiral algebra of the Sp(2$N$) vector model at large $N$, and more generally in the chiral algebra of the 't\;Hooft limit of the coset model. Note that, up to a central element, $D(2,1;\alpha)$ is isomorphic to the wedge algebra of the large $\mathcal{N}=4$ superconformal algebra, c.f. \cite{Gaberdiel:2013vva}. By matching the spectrum of massless gauge fields in the bulk AdS$_3$ theory with the chiral spectrum of a CFT$_2$, we have in this way performed the first check that the Vasiliev higher spin theory with gauge algebra $\mathfrak{shs}_2^{\mathfrak{sp}}[\mu]$ is dual to the Sp$(2N)$ coset model in the 't Hooft limit, whose chiral algebra realises $\mathcal{W}_{\infty}^{\text{e},\; \mathcal{N}=4}[\lambda]$. As in previous similar results, in the next section we will see that the different parameters are related as $\lambda=\mu$. For $\mu=0$ this relates the Vasiliev theory with the Sp($2N$) vector model.

%%%%%%%%%%%%%%%%%%%%%%%%%%%%%%%%%%%%%%%%%%%%%%%%%%%%%%%%%%%%%%%%
\subsection{Massive spectrum}\label{subsec:MASSIVE}
%%%%%%%%%%%%%%%%%%%%%%%%%%%%%%%%%%%%%%%%%%%%%%%%%%%%%%%%%%%%%%%%

Having matched the CFT chiral spectum with the spectrum of massless fields in AdS, to achieve full correspondence we also have to match representations of the CFT with (generically massive) matter degrees of freedom in the bulk. The fundamental representations of the $\mathfrak{shs}_2^{\text{sp}}[\mu]$ algebra are the same as those of $\mathfrak{shs}_2[\mu]$, up to a reality condition. These can be obtained from the two fundamental representations of $\mathfrak{shs}[\mu]$, as seen in \cite{Gaberdiel:2013vva}, which are constructed from two short representations of $D(2,1;\alpha)$, denoted $\phi_{\pm}$, with $L_0$-eigenvalues
\begin{alignat}{3}
  h_+=\frac{\mu}{2}, & \qquad &  h_-=\frac{1}{2}(1-\mu) \; .
\end{alignat}
The fundamental representations of $\mathfrak{shs}^{\text{sp}}_2[\mu]$ can be constructed from these by taking the tensor product
\begin{equation}
 \phi_{\pm}\otimes\mathbf{2} \; ,
\end{equation}
where $\mathbf{2}$ is the fundamental representation of the matrix algebra. In this way, there is a doublet of states with $L_0$ eigenvalues $h_{\pm}$, forming short supermultiplets. Their quantum numbers with respect to $\mathfrak{su}(2)_{\pm}$ are
\begin{alignat}{3}
 \phi_+:(\mathbf{2},\mathbf{1})\oplus(\mathbf{1},\mathbf{2}) &\qquad & \phi_-:(\mathbf{1},\mathbf{2})\oplus(\mathbf{2},\mathbf{1}) \; ,
\end{alignat}
where we have used that the supercharges transform as $({\bf 2},{\bf 2})$, and picked the anti-symmetric part of the tensor product.\\

As argued in \cite{Gaberdiel:2013vva}, these correspond to two massive scalars and two Dirac fermions propagating in AdS. Given these properties, we can identify the corresponding degrees of freedom in the coset CFT side as
\begin{alignat}{3}
 \phi_+\leftrightarrow (v;0) \; , & \qquad & \phi_-\leftrightarrow (0;v) \; .
\end{alignat}
Indeed, these are BPS states whose conformal dimensions are precisely $h_{\pm}$, if we take the 't Hooft limit and identify $\mu$ with the 't Hooft parameter $\lambda$, see \ref{app:BPS}. Furthermore, $\phi_{\pm}$ must be real scalars, since the fundamental representation $v$ on the coset side is self-conjugate. \\

With all this in mind, the one-loop partition function of the bulk theory, consisting of $\mathfrak{shs}^{\text{sp}}_2[\mu]$ and the real scalars above, can be matched with the 't Hooft limit partition function of the coset, with the identification $\lambda =\mu$, see Appendix \ref{app:DUAL}.

%%%%%%%%%%%%%%%%%%%%%%%%%%%%%%%%%%%%%%%%%%%%%%%
\subsection{Comments on finite $N$ effects}
%%%%%%%%%%%%%%%%%%%%%%%%%%%%%%%%%%%%%%%%%%%%%%%

It is not clear whereas $\mathcal{W}_\infty^{\text{e},\mathcal{N}=4}[\lambda]$ is a subalgebra of $\mathcal{W}_\infty^{\mathcal{N}=4}[\lambda]$.
This is the case if the quantum DS reduction of the higher spin algebra is shown to commute with the truncation automorphism.
This issue will not be analysed here.
In the quantum case, i.e. for finite $N$ and $k$ (and therefore finite central charge $c$), the model is not expected to be a mere truncation of the original $\mathcal{W}_{N,k}^{\mathcal{N}=4}[\lambda]$ construction \cite{Beccaria:2014jra}.\\
\\
In \cite{Candu:2012ne} it was found that there are two natural ways in which the free parameter $\gamma$ of the quantum bosonic even spin $\mathcal{W}_\infty$-algebra can be identified with $\lambda$ at finite $c$.
These two ways agree in the classical limit $c\rightarrow\infty$, and they correspond to two different quantisations of the classical DS reduction of the even spin bosonic algebra $\mathfrak{hs}^{\text{e}}[\lambda]$.
This was seen as a reflection of the fact that $\mathfrak{hs}^{\text{e}}[\mu]$ truncates for $\mu=N$ to either $\mathfrak{sp}(N)$ if $N$ is even, or $\mathfrak{so}(N)$ if $N$ is odd.
Note that these algebras are Langlands dual.
Just as in \cite{Candu:2013uya}, we expect that such ambiguities are also present for $\mathcal{W}_\infty^{\text{e},\mathcal{N}=4}[\lambda]$.\\

It is known that $\mathfrak{shs}[\mu=N]$ has an ideal $\chi_N$, such that $\mathfrak{shs}[\mu=N]/\chi_N = \mathfrak{sl}(N\vert N-1)$, see e.g. \cite{Candu:2013uya} and references therein.
For the extended higher spin algebras $\mathfrak{shs}_2[\mu]$, this ideal has an extended version.
Since $\mathfrak{sl}(2)$ is simple, its ideals are the null element $0_2$, and $\mathfrak{sl}(2)$ itself. 
Therefore $\chi_N\otimes 0_2$ is a non-trivial ideal of $\mathfrak{shs}_2[N]$, and its truncation under this ideal is
\begin{equation}
\mathfrak{shs}_2\left[\mu=N\right]/\chi_N\otimes 0_2 = \mathds{1}\otimes\mathfrak{psl}(2)\oplus \mathfrak{sl}(N\vert N-1)\otimes \mathds{1}_2\oplus \mathfrak{sl}(N\vert N-1)\otimes\mathfrak{psl}(2) \, ,
\end{equation}
where $\mathfrak{psl}(2)=\mathfrak{sl}(2)/0_2$.
We would like to know what is the effect of truncations by automorphisms on these algebras.
In \cite{Candu:2013uya} it was found that for odd $N$ the algebra truncates to $\mathfrak{osp}(N\vert N-1)$, whereas for even $N$ it reduces to $\mathfrak{osp}(N-1\vert N)$.
Writting $N=2n+1$ or $N=2n+2$ for the two distinct cases, we get $ B(n,n)$ and $B(n,n+1)$, respectively.
Note that $B(n,n)$ is Langlands self-dual, whereas $B(n,n+1)$ gets mapped to $B(n+1,n)$.
It seems therefore natural to conjecture that the chiral algebra of the coset \eqref{eqn:coset} at finite $N$, $k$, is the DS reduction of $B(n,n)\otimes\mathfrak{psl}(2)$ or $B(n+1,n)\otimes\mathfrak{psl}(2)$ (plus the terms with the identity elements), depending on the parity of $N$, as in \cite{Candu:2013uya}.
It is an open question whereas it is possible to find cosets whose chiral algebra at finite $N$ match the DS reduction of several other extended Lie superalgebras.
These DS reductions cannot have $\mathcal{N}=4$ rank, since the cosets with this amount of supersymmetry were listed in \cite{Sevrin:1989ce}.
Such analysis is beyond the scope of this paper, and will be studied elsewhere.

%%%%%%%%%%%%%%%%%%%%%%%%%%%%%%%%%%%%%%%%%%%%%%%%%%%%%
\section{Relation with the symmetric product}\label{sec:SYMM}
%%%%%%%%%%%%%%%%%%%%%%%%%%%%%%%%%%%%%%%%%%%%%%%%%%%%%

The proposed CFT$_2$ dual to string theory in AdS$_3$ $\times$ S$^3$ $\times$ $\mathbb{T}^4$ at the tensionless point (see \cite{David:2002wn,Gaberdiel:2014cha} and references therein) is given by the symmetric product of 4$(N+1)$ free bosons and fermions
\begin{equation}\label{eqn:SYMM}
\left( \mathbb{T}^4 \right)^{N+1}/S_{N+1} \; ,
\end{equation}
composed of $N+1$ copies of four free bosons and fermions, with the symmetric group acting on the copies. In the same way as in  \cite{Gaberdiel:2014cha,Gaberdiel:2015mra} for the U($N$) vector model, the untwisted sector of the Sp($2N$) vector model can be identified with a subsector of the untwisted sector of the symmetric product, since $S_{N+1}\subset$ Sp($2N$). This task is greatly simplified by the observation that the symmetric group $S_{N+1}$ is a subgroup of Sp(2$N$) via the group embeddings
\begin{equation}
 S_{N+1}\subset \text{U}(N) \subset \text{Sp}(2N) \; ,
\end{equation}
where $S_{N+1}\subset \text{U}(N)$ was constructed in \cite{Gaberdiel:2014cha}, and $\text{U}(N) \subset \text{Sp}(2N)$ can be found in Appendix \ref{app:coset}. 
By decomposing the untwisted sector of the U($N$) vector model into representations of the untwisted sector of the Sp($2N$) vector model, we can then use the results of \cite{Gaberdiel:2014cha} to decompose the untwisted sector of the symmetric product into representations of the untwisted sector of the Sp($2N$) vector model.\\

As a first check, note that the vector model contains $4(N+1)$ bosons and fermions transforming as
\begin{equation}
2\times(\mathbf{2N})\oplus 4\times(\mathbf{1})\; ,
\end{equation}
and under the embedding above the vector representation of Sp$(2N)$ splits as 
\begin{equation}                                                                                                                                                                           (\mathbf{2N})\rightarrow\mathbf{N}\oplus\mathbf{\bar{N}}\rightarrow 2\times (N) \; ,                                                                                                                                
\end{equation}
where ${\bf N}$, ${\bf \bar{N}}$ denote the fundamental and anti-fundamental representations of U($N$), respectively, and $(N)$ is the irreducible standard representation of $S_{N+1}$,. Under $S_{N+1}\subset$ Sp($2N)$, the transformation rules of the free bosons and fermions under $S_{N+1}$ are then
\begin{equation}
4\times (N) \oplus 4\times ({\bf 1}) = 4\times (N+1)\; , 
\end{equation}
where $(N+1)$ is the reducible representation of $S_{N+1}$ given by permutation matrices. This matches the transformation rules of the basic fermionic and bosonic constituents of the symmetric orbifold theory \eqref{eqn:SYMM}.\\

Note that since we will sit at $\mu=0$ throughout this section, we will only keep track of the quantum numbers with respect to $\mathfrak{su}(2)_+$.

%%%%%%%%%%%%%%%%%%%%%%%%%%%%%%%%%%%%%%%%%%%%%%%%%%%%%%%%%%
\subsection{Decomposing the U($N$) untwisted sector}
%%%%%%%%%%%%%%%%%%%%%%%%%%%%%%%%%%%%%%%%%%%%%%%%%%%%%%%%%%

Throughout the following sections we will extensively use the coset notation for the vector models. To settle notation, recall that the representations in the untwisted sector of the vector model are described by the $k\rightarrow\infty$ limit of the coset representation $(0;\Xi)$, where $\Xi$ is a representation of Sp($2N$), obtained from the vector representation ${\bf 2N}\equiv v$ by taking successive tensor products. The same holds for the U($N$) orbifold of \cite{Gaberdiel:2014cha,Gaberdiel:2015mra}, whose representations in the untwisted sector were given by the $k\rightarrow\infty$ limit of U($N$)-coset representations $(0;\Lambda)$, where $\Lambda$ is a representation of U($N$), obtained from the fundamental $f\equiv{\bf N}$ and anti-fundamental $\bar{f}\equiv{\bf \bar{N}}$ representations by taking successive tensor products.\\

The characters of the coset representations $(0;\Xi)$ are denoted as $\chi^{\text{sp}}_{(0;\Xi)}(q,y)$, where $\Xi$ is a representation of Sp($2N$), whereas $q$ and $y$ keep track of the conformal dimension and $\mathfrak{su}(2)_+$ chemical potential, respectively. The representation $\Xi$ is labelled by $N$ Dynkin labels $\Xi_i$, and we will use the notation
\begin{equation}\label{eqn:decompos}
\Xi \equiv \langle \Xi_1,\Xi_2,\ldots, \Xi_{N}\rangle \; ,
\end{equation}
for Sp($2N$) representations. The characters of the $(0;\Lambda)$ representations of the U$(N)$-coset are denoted $\chi_{(0;\Lambda)}(q,y)$, where $\Lambda$ is a U($N$) representation. Under U$(N)\subset$ Sp($2N$) they decompose as
\begin{equation}
 \chi_{(0;\Lambda)}(q;y)=\sum_{\Xi}n(\Lambda;\Xi)\chi^{\text{sp}}_{(0;\Xi)}(q;y) \; ,
\end{equation}
where $n(\Lambda;\Xi)$ is the multiplicity of $\Lambda$ in $\Xi$. For $\Lambda =0$ the decomposition can be found to be
\begin{equation}\label{eqn:VAC_DEC}
 \chi_{(0;0)}(q,y) = \sum_{r,s} \chi^{\text{sp}}_{(0;\Xi_{r,s})} \; ,
\end{equation}
where $\Xi_{r,s}$ are the Sp$(2N)$ representations
\begin{alignat}{3}\label{eqn:DEC}
 \Xi_{r,s} \equiv \langle 0,\ldots 0,r,0, \ldots ,0,s,0, \ldots,0\rangle \; ,& \qquad & r,s\in 2\mathbb{N}^0 \; ,
\end{alignat}
with $r$, $s$ sitting at any two positions.\\

We would like to check \eqref{eqn:VAC_DEC} by explicitly constructing the corresponding states in the vector models. For $k\rightarrow\infty$ and sufficiently large $N$ \cite{Gaberdiel:2014cha}, the characters can be written as
\begin{align}\label{eqn:wedge}
\begin{aligned}
 \chi_{(0;\Lambda)}(q,y) & =  \chi^{(\text{wedge})}_{(0;\Lambda)}(q,y)\cdot \chi_0(q,y) \\
  \chi^{\text{sp}}_{(0;\Xi)}(q,y) & =  \chi^{(\text{wedge})}_{(0;\Xi)}(q,y)\cdot \chi_0^{\text{sp}}(q,y) \; , 
  \end{aligned}
\end{align} 
where $\chi^{(\text{wedge})}_{(0;\Lambda)}$ is the character with respect to the corresponding wedge algebra, and $\chi_0(q,y)$, $\chi_0^{\text{sp}}(q,y)$ are the characters of the chiral algebra of the U$(N)$ (given in \cite{Gaberdiel:2013vva}), and Sp($2N$) (given in \eqref{eqn:VAC}) vector models, respectively. Equations \eqref{eqn:wedge} enable us to rewrite \eqref{eqn:VAC_DEC} as
\begin{equation}\label{eqn:RS}
 \frac{\chi_0(q;y)}{\chi_0^{\text{sp}}(q;y)}=\sum_{r,s}\chi^{(\text{wedge})}_{(0;\Xi_{r,s})} \; ,
\end{equation}
where we used $\chi^{(\text{wedge})}_{(0;0)}(q,y)=1$.\\

In order to check \eqref{eqn:RS}, we need the wedge characters. As suggested by the notation, we claim that the wedge characters of the Sp($2N$) vector model are the same as those for the U($N$) vector model, with the $\Xi_i$ interpreted as U($N$) Dynkin labels. We give a brief argument in favour of this claim in appendix \ref{app:NULL}.\\

Substituting for the expressions of the vacuum characters, equation \eqref{eqn:RS} becomes
\begin{align}\label{eqn:RATIO}
\begin{aligned}
\sum_{r,s}\chi^{\text{sp},(\text{wedge})}_{(0;\Xi_{r,s})}= &  \prod_{s=1}^{\infty}\prod_{n=s}^{\infty}(1+y^{1/2}q^{n+1/2})^2(1+y^{-1/2}q^{n+1/2})^2\\
 & \times\prod_{s\text{ even}}\prod_{n=s}^{\infty}\frac{1}{(1-q^n)^4(1-yq^n)(1-y^{-1}q^n)}\\
& \times\prod_{\substack{s\geq 3\\s\text{ odd}}}\prod_{n=s}^{\infty}\frac{1}{(1-q^n)^2}\\
& \times \prod_{n=1}^{\infty}\frac{1}{1-q^n} \; .
\end{aligned}
\end{align}
Expanding the right-hand side to $\mathcal{O}(q^{3})$,
 \begin{align}\label{eqn:RHS}
 \begin{aligned}
1 & +q+2q^{3/2}(y^{1/2}+y^{-1/2})\\
 & +q^2(y^{-1}+y+6)+6q^{5/2}(y^{1/2}+y^{-1/2})+\mathcal{O}(q^{3}) \; ,
 \end{aligned}
 \end{align}
we can check explicitly that it equals
\begin{align}
  1+\chi_{(0;\langle 2,0,\ldots,0\rangle)}^{(\text{wedge})}+\chi_{(0;\langle 0,2,0,\ldots,0\rangle)}^{(\text{wedge})}+\chi_{(0;\langle 0,0,2,0\ldots,0\rangle)}^{(\text{wedge})}+\chi_{(0;\langle 4,0,\ldots,0\rangle)}^{(\text{wedge})} \; ,
 \end{align}
up to order $q^{5/2}$, using the ancillary file of \cite{Gaberdiel:2014cha} for the explicit form of the wedge characters.\\

Microscopically, the expression \eqref{eqn:RHS} counts the bilinears which are U$(N)$ singlets but not Sp$(2N)$ singlets. At $h=1$ this is given by (see Appendix \ref{app:coset} for the conventions and notations) 
\begin{equation}\label{eqn:BILIN}
 \sum_{i=1}^N \left( \psi_{-1/2}^{i,\alpha}\psi_{-1/2}^{-i,\beta} + \psi_{-1/2}^{-i,\alpha}\psi_{-1/2}^{i,\beta}\right)\vert 0\rangle  \; ,
\end{equation}
which is a singlet under the R-symmetry since we pick the antisymmetric product of ${\bf 2}\otimes{\bf 2}$. Note that the relative sign between the two terms in \eqref{eqn:BILIN} ensures that this is not an Sp$(2N)$ singlet, whereas taking different signs for the Sp$(2N)$ labels makes each term a U$(N)$ singlet. At $h=3/2$ we have
\begin{equation}\label{eqn:H32}
 \sum_{i=1}^N \left( \psi_{-1/2}^{i,\alpha}\mathcal{J}_{-1}^{-i,\beta}+\psi_{-1/2}^{-i,\alpha}\mathcal{J}_{-1}^{i,\beta} \right)\vert 0\rangle  \; ,
\end{equation}
for $\beta=\pm$, which transforms in the $\mathbf{2}$ of $\mathfrak{su}(2)_+$. At $h=2$ there are four singlets coming from
\begin{equation}\label{eqn:JBILIN}
 \sum_{i=1}^N \left( \mathcal{J}_{-1}^{i,\alpha}\mathcal{J}_{-1}^{-i,\beta} +\mathcal{J}_{-1}^{-i,\alpha}\mathcal{J}_{-1}^{i,\beta}\right) \vert 0\rangle \; ,
\end{equation}
together with a triplet and a singlet from
\begin{equation}\label{eqn:tripletH2}
 \sum_{i=1}^N \left( \psi_{-3/2}^{i,\alpha}\psi_{-1/2}^{-i,\beta} + \psi_{-3/2}^{-i,\alpha}\psi_{-1/2}^{i,\beta}\right) \vert 0\rangle \; .
\end{equation}
These are easily seen to agree with the first few terms of \eqref{eqn:RHS}.

%%%%%%%%%%%%%%%%%%%%%%%%%%%%%%%%%%%%%%%%%%%%%%%%%%%%
\subsection{Decomposing the $S_{N+1}$ untwisted sector}
%%%%%%%%%%%%%%%%%%%%%%%%%%%%%%%%%%%%%%%%%%%%%%%%%%%%

The decomposition of the single particle symmetry generators of the untwisted sector of the symmetric orbifold into representations of the untwisted sector of the U$(N)$ vector model was given in \cite{Gaberdiel:2015mra} as
\begin{align}\label{eqn:UN}
\begin{aligned}
 \sum_{r,l}N(r,l)q^ry^l= & \left(y^{1/2}+y^{-1/2}\right)\left[2q^{1/2}+\frac{4q^{3/2}}{1-q}\right]+\sum_{n=1}^{\infty}\left(y+6+y^{-1}\right)q^n \\
& +\left(1-q\right)\sum_{m,n\geq 0}{}^{'}\chi_{(0;[m,0,\ldots,0,n])}^{(\text{wedge})\mathcal{N}=4[0]}(q,y) \; ,
\end{aligned}
 \end{align}
where $\sum^{'}$ denotes the fact that the terms $(m,n)=(1,0),(0,1),(1,1)$ are excluded from the sum. These terms appear isolated in the first line of \eqref{eqn:UN}, and they correspond to the generators of $\mathcal{W}_{\infty}^{\mathcal{N}=4}[0]$. The single particle generators correspond then to the representations of the U($N$) vector model given by
 \begin{equation}
(0;[m,0,\ldots,0,n]) \; ,
 \end{equation}
that is, the $m$-th tensor power of the ${\bf N}$ bosons and ${\bf N}$ fermions, and the $n$-th tensor power of the ${\bf \bar{N}}$ bosons and ${\bf \bar{N}}$ fermions. In order to obtain a similar decomposition in terms of Sp$(2N)$ representations, we note that
 \begin{align}
  \nonumber \chi_{(0;[m,0,\ldots,0,n])}^{(\text{wedge})} & =  \chi_{(0;[m,0,\ldots,0,0])}^{(\text{wedge})}\cdot   \chi_{(0;[0,0,\ldots,0,n])}^{(\text{wedge})}\\
   & =   \chi_{(0;[m,0,\ldots,0,0])}^{(\text{wedge})}\cdot  \chi_{(0;[n,0,\ldots,0,0])}^{(\text{wedge})}\\
  \nonumber & =  \sum_{\substack{r,s\\r+2s=m+n}}\chi_{(0;[r,s,0,\ldots,0])}^{(\text{wedge})} \; ,
 \end{align}
where we have used the decomposition 
\begin{equation}
 [m,0,\ldots,0]\otimes [n,0,\ldots,0]=\sum_{\substack{r,s\\r+2s=m+n}}[r,s,0,\ldots,0] \; ,
\end{equation}
also valid as a fusion rule for wedge representations of the U($N)$ vector model.\\

The case $(m,n)=(1,1)$ is of particular importance since it corresponds to some of the generators of $\mathcal{W}_{\infty}^{\mathcal{N}=4}[0]$:
 \begin{equation}\label{eqn:DECOMP}
   \chi_{(0;[1,0,\ldots,0,1])}^{(\text{wedge})} = \chi_{(0;[2,0,\ldots,0,0])}^{(\text{wedge})} + \chi_{(0;[0,1,0,\ldots,0,0])}^{(\text{wedge})} \; .
 \end{equation}
Observe now that
\begin{align}
\begin{aligned}
  \left(1-q\right)\left[ 2\chi_{(0;[1,0,\ldots,0])}^{(\text{wedge})}+ \chi_{(0;[0,1,0,\ldots,0])}^{(\text{wedge})}\right] & =
\left(y^{1/2}+y^{-1/2}\right)\;\frac{2q^{1/2}}{1-q}+\left(y+y^{-1}+5\right)q^1  \\
&   + \sum_{n\text{ odd}}\left(y+y^{-1}+4\right)q^n+2\sum_{n\text{ even}}q^n \; ,
\end{aligned}
\end{align}
accounts precisely for the generators of $\mathcal{W}_{\infty}^{\text{e},\; \mathcal{N}=4}[0]$, which is the chiral algebra of the Sp$(2N)$ vector model. The fact that the representation $(0;[2,0,\ldots,0])$ corresponds to generators of $\mathcal{W}_{\infty}^{\mathcal{N}=4}[0]$ but not of $\mathcal{W}_{\infty}^{e,\; \mathcal{N}=4}[0]$ is simply a consequence of \eqref{eqn:RS} and \eqref{eqn:DEC}.\\

We are now able to decompose the character of the untwisted sector of the symmetric orbifold in terms of Sp$(2N)$ characters:
\begin{align}\label{SP}
\begin{aligned}
 \sum_{r,l}N(r,l)q^ry^l & =  2\left(y^{1/2}+y^{-1/2}\right)\frac{q^{1/2}}{1-q}+\sum_{n\text{ odd}}^{\infty}\left(y+4+y^{-1}\right)q^n+2\sum_{n\text{ even}}q^n\\
& +\left(y+y^{-1}+5\right)q+\left(1-q\right)\sum_{n,m\geq 0}{}^{'}\left(n+1\right)\cdot\chi_{(0;\langle n,m,0,\ldots,0\rangle)}^{(\text{wedge})}(q,y) \; ,
\end{aligned}
 \end{align}
where the prime in the sum indicates that the cases $(n,m)=(1,0),(0,1)$ are excluded (these are precisely the terms written explicitly). The single particle generators of the untwisted sector of the symmetric orbifold correspond then to the representations of the Sp($2N$) vector model given by
\begin{equation}\label{eqn:sing_part}
(n+1)\cdot(0;\langle n,m,0,\ldots ,0\rangle) \; ,
\end{equation}
with the cases $(n,m)=(1,0)$, $(n,m)=(0,1)$ corresponding to the generators of $\mathcal{W}_{\infty}^{\text{e},\; \mathcal{N}=4}[0]$ itself.\\

The higher spin square is now constructed from 4$N$ \emph{real} free fermions and bosons, instead of 2$N$ \emph{complex} fields, as in the original formulation \cite{Gaberdiel:2014cha,Gaberdiel:2015mra,Gaberdiel:2015wpo,Gaberdiel:2017ede}.
The degeneracies in \eqref{eqn:sing_part} account for the multiplicity of the corresponding U($N)$ representation in a Sp($2N)$ representation -- this was denoted $n(\Lambda;\Xi)$ in \eqref{eqn:decompos}.
Then, contrary to the original higher spin square construction, we will have more than one field per site in the square: columns are labelled by the number of boxes of the corresponding diagram $n+2m$, and each column comes with the multiplicity $n+1$.\\

For simplicity we restrict our attention to $2N$ free fields $\phi^{\alpha i}$, where $\alpha=\pm$ is an (auxiliary) SU(2) fundamental label, and $i=1,\ldots, N$ is a label of the irreducible standard representation of $S_{N+1}$.
The single particle generators \eqref{eqn:sing_part} can then be constructed from these fundamental free fields by using the following rule: the only allowed fields are the ones which are totally symmetric under $S_{N+1}\otimes$ SU(2).
Explicitly, we find the correct multiplicities by starting with a fundamental field $\yng(1)_{~{\bf 2}}$ and symmetrising its self-products on both labels.
We denote this fusion rule by the symbol $\otimes_{\text{s}}$.\\

It is instructive to check this for the first few Young diagrams, labelled by the number of boxes $n+2m$.
For $n+2m=1$ we have $(n,m)=(1,0)$, which corresponds to the sum over the $S_{N+1}$ label of the fundamental fields 
\begin{alignat}{4}
n+2m=1: & \qquad & \yng(1)_{~{\bf 2}} & \, .
\end{alignat}
For two boxes $n+2m=2$, we have $(n,m)=(2,0)$ or $(n,m)=(0,1)$, which arise with the multiplicities:
\begin{alignat}{4}
n+2m=2: & \qquad & \yng(1)_{~{\bf 2}}\, \otimes_\text{s} & \, \yng(1)_{~{\bf 2}}= \yng(2)_{~{\bf 3}}\oplus \yng(1,1)_{~{\bf 1}} \, .
\end{alignat}
Here we have either anti-symmetrised or symmetrised on both indices simultaneously, in order to keep the total product symmetric.
For three boxes, we get
\begin{alignat}{3}
\begin{aligned}
n+2m=3: & \qquad & \yng(2)_{~{\bf 3}}\, \otimes_\text{s} & \,  \yng(1)_{~{\bf 2}}= \yng(3)_{~{\bf 4}}\oplus \yng(2,1)_{~{\bf 2}}\\
		& \qquad & \yng(1,1)_{~{\bf 1}}\, \otimes_\text{s} & \, \yng(1)_{~{\bf 2}}= \yng(2,1)_{~{\bf 2}} \, .
\end{aligned}
\end{alignat}
Note that the last diagram of the first line is the same as the diagram in the second line. 
For four boxes, we get
\begin{alignat}{3}
\begin{aligned}
n+2m=4: & \qquad & \yng(3)_{~{\bf 4}}\, \otimes_\text{s} & \,  \yng(1)_{~{\bf 2}}= \yng(4)_{~{\bf 5}}\oplus \yng(3,1)_{~{\bf 3}}\\
		& \qquad & \yng(2,1)_{~{\bf 2}}\, \otimes_\text{s} & \, \yng(1)_{~{\bf 2}}= \yng(3,1)_{~{\bf 3}}\oplus \yng(2,2)_{~{\bf 1}} \, .
\end{aligned}
\end{alignat}
Again, note that the last diagram of the first line and the first diagram of the second line denote the same physical field.
These multiplicities match those of \eqref{eqn:sing_part} up to this order.

%%%%%%%%%%%%%%%%%%%%%%%%%%%%%%%%%%%%%%%%%%%%%%%%%%%%%%%%%%
\subsection{The symmetric product of $K3$}
%%%%%%%%%%%%%%%%%%%%%%%%%%%%%%%%%%%%%%%%%%%%%%%%%%%%%%%%%%

Note that the results of \cite{Baggio:2015jxa}, regarding the symmetric product of K3 at the point $K3\cong\mathbb{T}^4/\mathbb{Z}_2$, can also be expressed in terms of the representations of the untwisted sector of the Sp($2N$) vector model. It is argued in \cite{Baggio:2015jxa} that instead of $\mathcal{W}^{\mathcal{N}=4}_{\infty}[0]$ one must consider the subalgebra obtained by removing the four singlet bosons and fermions. This is due to the fact that $\mathbb{Z}_2$ acts by exchanging the sign of the fundamental fields, so that all states of the vector model built using an odd number of fields are projected out. The bilinear basic invariants remain, but the singlet fields are left out. Using the notation of \cite{Baggio:2015jxa}, the decomposition of the single particle generators of the untwisted sector of the symmetric product of K3 in terms of Sp($2N$) representations can be found to be
\begin{align}
J^{\text{K3}}(q,y) & = \frac{2q^{3/2}}{1-q}\left(y^{1/2}+y^{-1/2}\right)+\left(y+y^{-1}+1\right)q+\sum_{n\text{ odd}}^{\infty}\left(y+4+y^{-1}\right)q^n+2\sum_{n\text{ even}}q^n  \nonumber \\
& +\left(1-q\right)\sum_{\substack{n,m\geq 0\\n\text{ even}}}{}^{'}\left(n+1\right)\cdot\chi_{(0;\langle n,m,0,\ldots,0 \rangle)}^{(\text{wedge})}(q,y) \; ,
\end{align}
where again the prime in the sum indicates that the cases $(n,m)=(1,0),(0,1)$ are excluded.

%%%%%%%%%%%%%%%%%%%%%%%%%%%%%%%%%
\section{Conclusion}
%%%%%%%%%%%%%%%%%%%%%%%%%%%%%%%%%

In this paper we proposed a new higher spin/CFT duality, between $\mathcal{N}=4$ theories with a spectrum of superprimary fields with even spin. The full correspondence was obtained by adding to the bulk theory generally massive real scalar fields and their fermionic superpartners, corresponding to representations of the chiral algebra of the dual CFT. Several checks of this duality were performed, including the matching of the symmetry generating spectrum, and the one-loop partition functions. Both sides of the duality are deformable, while preserving their symmetries, and give rise to a 1-parameter family of theories which are dual to each other.\\

When this parameter is tuned to vanish, the dual CFT becomes a symplectic vector model of free bosons and fermions, which can be studied in great detail. Furthermore, at this point it is possible to embed the untwisted sector of the vector model in the untwisted sector of the symmetric product theory, believed to be dual to string theory in AdS$_3\;\times\;\text{S}^3\;\times\mathbb{T}^4$ at the tensionless point. The details of this embedding for the symplectic model were worked out in detail, and give rise to an alternative description of the stringy symmetries.\\

Further work is required in order to establish the existence and uniqueness of the even spin $\mathcal{N}=4$ $\mathcal{W}_{\infty}$-algebra realised by the coset. As in \cite{Gaberdiel:2014yla}, the study of the asymptotic symmetry algebra of $\mathfrak{shs}_2^{\text{sp}}[\mu]$ and its matching with the 't Hooft limit of the coset algebra would constitute a strong check of the proposed holographic duality. 
In the same way, the study of the different algebras and their matching at finite $N$, $k$ would constitute a strong argument in favour of the proposed duality.
It would also be interesting to study other extended truncated Vasiliev theories with $\mathcal{N}=4$. In particular, the $\mathfrak{shs}_4[\mu]$ theory has $\mathcal{N}=6$ rank, which consistently reduces to $\mathcal{N}=4$ after an SO-like truncation.
Nevertheless, this theory is not dual to the SO-type coset of \cite{Sevrin:1989ce}: due to the problematic issues of this coset presented in the introduction, their spectrum does not agree.
In other words, using the standard boundary conditions, the asymptotic symmetry algebra does not preserve the $\mathcal{N}=4$ symmetry of the bulk theory.
The precise mechanism responsible for this, as well as its relation to string theory, is going to be analysed elsewhere. 

%%%%%%%%%%%%%%%%%%%%%%%%%%%%%%%%%%%%%%%%%%%%%%%%%%%%%%%%%%%%%%%%%%%%%
\acknowledgments
%%%%%%%%%%%%%%%%%%%%%%%%%%%%%%%%%%%%%%%%%%%%%%%%%%%%%%%%%%%%%%%%%%%
It is a pleasure to thank Matthias Gaberdiel for guidance, help, and numerous discussions throughout the realisation of this work, which were crucial for its successful completion. We also thank Juan Jottar for a careful reading of the manuscript and very valuable advice, as well as Shouvik Datta, Arvin Moghaddam, and Cheng Peng for various very helpful discussions. Finally, we thank Carl Vollenweider and Constantin Candu, whose notes on previous work were of great help.

%%%%%%%%%%%%%%%%%%%
\appendix
%%%%%%%%%%%%%%%%%%%
%%%%%%%%%%%%%%%%%%%%%%%%%%%%%%%%%%%%%%%%%%%%%%%%%%%%%%%%%%%%%%%%
\section{Chiral algebra of the vector model}\label{subsec:VAC}
%%%%%%%%%%%%%%%%%%%%%%%%%%%%%%%%%%%%%%%%%%%%%%%%%%%%%%%%%%%%%%%%

In this section we present a detailed derivation of the vacuum character of the $\mathcal{N}=4$ Sp(2$N$) vector model. We mainly follow the methods of \cite{Candu:2012jq,Candu:2014yva}.\\

The chiral algebra of the vector model at large $N$ is given by all the Sp($2N$)-invariant combinations of the bosonic and fermionic fields in \eqref{eqn:REPSR} and \eqref{eqn:REPSR2}. In order to find its character, we denote the two copies of the free vector bosonic currents transforming in the ${\bf 2N}$ of Sp(2$N$) as $\mathcal{J}^{i,\alpha}$, with $i=\pm 1,\ldots,\pm N$ (see Appendix \ref{app:coset} for conventions), and $\alpha=\pm$ labels the states in a doublet of $\mathfrak{su}(2)_-$. The four bosonic currents in the singlet of Sp$(2N)$ are denoted as $X^l$, $l=1,2,3$ an adjoint label of $\mathfrak{su}(2)_-$, together with $X^4$, which in uncharged under the R-symmetry. With the same conventions for the indices, the free fermionic vector NS currents are denoted as $\psi^{i,\alpha}$, together with $\lambda^{\alpha\beta}$, with $\alpha$, $\beta = \pm$, labelling doublets of $\mathfrak{su}(2)_+$ and $\mathfrak{su}(2)_-$, respectively.\\

A straightforward set of Sp($2N$)-invariant states is given by all the combinations of the four singlet bosons and fermions:
\begin{equation}
\left( X^{p}_{-n-1} \right)^A\left( \lambda^{\alpha\beta}_{-r-1/2}\right)^B  \; ,
\end{equation}
with $n,r\in\mathbb{N}^0$, for $p=1,2,3,4$, $A\in\mathbb{N}^0$, $B=0,1$, and $\alpha, \beta=\pm$. The counting of all such states goes as usual for free fields. Defining the chemical potentials $y_{\pm}=e^{2\pi i J_0^{3,\pm}}$, where $J_0^{3,\pm}$ are the Cartan generators of $\mathfrak{su}(2)_{\pm}$, their contribution to the character is
\begin{equation}
\prod_{n=1}^{\infty}\frac{(1+y_+^{1/2}q^{n-1/2})(1+y_+^{-1/2}q^{n-1/2})(1+y_-^{1/2}q^{n-1/2})(1+y_-^{-1/2}q^{n-1/2})}{(1-y_-q^n)(1-y_-^{-1}q^n)(1-q^n)^2}\; .
\end{equation}

A more interesting contribution is obtained from the $2\times ({\bf 2N})$ free bosons and fermions. It is given by linear combinations of the basic invariants
\begin{equation}\label{eqn:state}
\left(\Omega_{ij}\psi^{i,\alpha_1}_{-r_1-\frac{1}{2}}\psi^{j,\beta_2}_{-s_1-\frac{1}{2}}\right)^{K(r_1,s_1)} \left(\Omega_{ij}\psi^{i,\alpha_2}_{-r_2-\frac{1}{2}}\mathcal{J}^{j,\beta_2}_{-n_2-1}\right)^{L(r_2,n_2)} \left(\Omega_{ij}\mathcal{J}^{i,\alpha_3}_{-n_3-1}\mathcal{J}^{j,\beta_3}_{-m_3-1}\right)^{M(n_3,m_3)} \; ,
\end{equation}
where $\Omega$ is the symplectic matrix in $2N$ dimensions
\begin{equation}
\Omega = \left( \begin{array}{cc}
0_N & \mathds{1}_N \\
-\mathds{1}_N & 0_N \\
\end{array}\right) \; ,
\end{equation}
with $0_N$, $\mathds{1}_N $ denoting the zero and identity matrices in $N$ dimensions, respectively. Note also that $K(r_1,s_1),M(n_3,m_3)\in\mathbb{N}^0$, whereas $L(r_2,n_2)=0,1$, for fixed $r_1,r_2,s_1,n_2,n_3,m_3\in\mathbb{N}^0$. Due to the symplectic nature of $\Omega$, some care is needed when counting the number of independent primaries. We start with the case $\alpha_i\neq \beta_i$, which amounts to two possibilities for the middle term in \eqref{eqn:state}, since we are contracting two different fields, and a single possibility for each of the other two, since for those we contract fields which transform in the same representation.\footnote{Note that this way of proceeding splits the fields artificially from the point of view of the representations of $\mathfrak{su}(2)_{\pm}$, but we can easily recover them when combining the different contributions.} For fixed $r_1,r_2,s_1,n_2,n_3,m_3$ the contribution of \eqref{eqn:state} is
\begin{equation}\label{eqn:KLM}
\left( q^{r_1+s_1+1} \right)^{K(r_1,s_1)} \left( q^{r_2+n_2+3/2} \right)^{L(r_2,n_2)}\left( q^{n_3+m_3+2}\right)^{K(n_3,m_3)} \; ,
\end{equation}
which summing over all possible $K(r_1,s_1),L(r_2,n_2),M(n_3,m_3)$ in this case leads to
\begin{equation}\label{eqn:KLM}
\prod_{r_1,s_1=0}^{\infty}\frac{1}{1-q^{r_1+s_1+1}}\prod_{r_2,n_2=0}^{\infty}\left(1+q^{r_2+n_2+3/2}\right)^2\prod_{n_3,m_3=0}^{\infty}\frac{1}{1-q^{n_3+m_3+2}} \; ,
\end{equation}
where the exponent in the middle term corresponds to the two different possibilities of taking $\alpha_2,\beta_2$. Defining different indices $n$ and $s$ for each one of the terms as $r_1+s_1+1=n$, $s=s_1+1$ for the first, $r_2+n_2+2=n$, $s=n_2+2$ for the second, and $n_3+m_3+2=n$, $s=m_3+2$ for the third, the conditions $r_1,r_2,n_3\geq 0$ become $n\geq s$, whereas $s_1,n_2,m_3\geq 0$ correspond to $s\geq 1$ in the first term, and $s\geq 2$ in the last two. Then \eqref{eqn:KLM} becomes
\begin{equation}
\prod_{n=1}^{\infty}\frac{1}{1-q^n}\prod_{s=2}^{\infty}\prod_{n=s}^{\infty}\frac{(1+q^{n-1/2})^2}{(1-q^n)^2} \; .
\end{equation}
\\
For the case $\alpha_i=\beta_i=\pm$, the basic invariant is still given by \eqref{eqn:state}, but after summing over $K,L,M$ we now get
\begin{equation}
\prod_{\substack{r_1,s_1=0\\r_1\leq s_1}}^{\infty}\frac{1}{1-q^{r_1+s_1+1}}\prod_{r_2,n_2=0}^{\infty}\left(1+q^{r_2+n_2+3/2}\right)^2\prod_{\substack{n_3,m_3=0\\n_3<m_3}}^{\infty}\frac{1}{1-q^{n_3+m_3+2}} \; ,
\end{equation}
where the conditions on $r_1,s_1$, and $n_3,m_3$ were introduced to avoid double counting, as well as counting of the combinations which are identically null, e.g. $\Omega_{ij}j_{-n-1}^ij^j_{-n-1}\equiv 0$ due to the antisymmetry of $\Omega_{ij}$. Proceeding as before, but with the different bounds for the indices, we obtain
\begin{equation}
\prod_{n=1}^{\infty}\frac{1}{(1-q^n)^2}\prod_{\substack{s\geq 3\\s\text{ odd}}}\prod_{n=s}^{\infty}\frac{1}{(1-q^n)^4}\prod_{s=2}^{\infty}\prod_{n=s}^{\infty}(1+q^{n-1/2})^2 \; ,
\end{equation}
where the squares come from the liberty of taking $\alpha_i=\beta_i=\pm$. Altogether, including also the states constructed from the $X^l$ and $\lambda^{\alpha\beta}$, the total number of states of the untwisted chiral sector of the vector model is
\begin{equation}\label{eqn:GEN}
\prod_{n=1}^{\infty}\frac{(1+q^{n-1/2})^4}{(1-q^n)^7}\prod_{\substack{s\geq 2\\s\text{ even}}}\prod_{n=s}^{\infty}\frac{1}{(1-q^n)^2}\prod_{\substack{s\geq 3\\s\text{ odd}}}\prod_{n=s}^{\infty}\frac{1}{(1-q^n)^6}\prod_{s=2}^{\infty}\prod_{n=s}^{\infty}(1+q^{n-1/2})^4.
\end{equation}
\\
Tracing back the R-symmetry quantum numbers of \eqref{eqn:REPSR}, \eqref{eqn:REPSR2} in \eqref{eqn:state}, we rewrite \eqref{eqn:GEN} as
\begin{align}\label{eqn:VAC}
& \prod_{n=1}^{\infty}\frac{(1+y_+^{1/2}q^{n-1/2})(1+y_+^{-1/2}q^{n-1/2})(1+y_-^{1/2}q^{n-1/2})(1+y_-^{-1/2}q^{n-1/2})}{(1-q^n)^3(1-y_-q^n)(1-y_-^{-1} q^n)(1-y_+q^n)(1-y_+^{-1}q^n)} \nonumber \\
\times & \prod_{\substack{s\geq 2\\s\text{ even}}}\prod_{n=s}^{\infty}\frac{1}{(1-q^n)^2}\prod_{\substack{s\geq 3\\s\text{ odd}}}\prod_{n=s}^{\infty}\frac{1}{(1-y_+q^n)(1-y_+^{-1}q^n)(1-y_-q^n)(1-y_-^{-1}q^n)(1-q^n)^2} \\
\times & \prod_{s=2}^{\infty}\prod_{n=s}^{\infty}(1+y_+^{1/2}q^{n-1/2})(1+y_+^{-1/2}q^{n-1/2})(1+y_-^{1/2}q^{n-1/2})(1+y_-^{-1/2}q^{n-1/2}) \; . \nonumber
\end{align}
\\
The first line accounts for the $\mathfrak{su}(2)_{\pm}$ adjoint R-currents and $\mathfrak{u}(1)$ current of $\mathcal{N}=4$ at spin 1, and the $({\bf 2},{\bf 2})$ spin-$1/2$ fermions $\lambda^{\alpha\beta}$, which are primaries, as well as their descendants. The subsequent lines correspond to four $({\bf 2},{\bf 2})$ fields of half-integer spin for $s\geq 3/2$, six fields $({\bf 3},{\bf 1})\oplus ({\bf 1},{\bf 3})$ of odd spin for $s>1$, and two $({\bf 1},{\bf 1})$ fields of every even spin.

%%%%%%%%%%%%%%%%%%%%%%%%%%%%%%%%%%%%%%%%%%%
\section{Coset model}\label{app:coset}
%%%%%%%%%%%%%%%%%%%%%%%%%%%%%%%%%%%%%%%%%%%

In this section we gather various useful conventions and technicalities concerning the relevant coset theory.\\

The coset theory
\begin{equation}\label{eqn:COSET3}
\frac{\mathfrak{sp}(2N+2)_{k+N+2}^{(1)}}{\mathfrak{sp}(2N)_{k+N+2}^{(1)}}\oplus\mathfrak{u}(1)^{(1)} \; ,
\end{equation}
was shown to have $\mathcal{N}=4$ in \cite{Sevrin:1989ce}. Here, $\mathfrak{g}^{(1)}_k$ denotes the supersymmetric Ka$\check{\text{c}}$-Moody algebra at level $k$, generated by the adjoint currents $J^a$, and their conformal weight 1/2 superpartners $\psi^a$, with $a=1,\ldots ,\text{dim}\; \mathfrak{g}$, whose modes satisfy
\begin{alignat}{3}
\left[ J^a_m, J^b_n \right] & = if^{ab}_{~~c}J^c_{m+n} + km\delta_{m,-n}\eta^{ab} \\
\left[ J^a_m, \psi^b_r \right] & = if^{ab}_{~~c}\psi^c_{m+r} \\
\left\lbrace \psi^a_r, \psi^b_s \right\rbrace & = k\eta^{ab}\delta_{r,-s} \; ,
\end{alignat}
where $f^{ab}_{~~c}$ are the structure constants of $\mathfrak{g}$, and $\eta^{ab}$ its Killing metric. The bosonic and fermionic currents can be decoupled by defining the currents
\begin{equation}
\mathcal{J}^a = J^a + \frac{i}{2k} f^a_{~bc}(\psi^b\psi^c) \; ,
\end{equation}
which satisfy a Ka$\check{\text{c}}$-Moody algebra at level $k-\check{h}$, where $\check{h}$ is the dual Coxeter number of $\mathfrak{g}$, as well as 
\begin{equation}
\left[ \mathcal{J}^a_m, \psi^b_r \right] = 0 \; ,
\end{equation}
and the fermions become manifestly free. In this way, the coset theory \eqref{eqn:COSET} is equivalent to
\begin{equation}
\frac{\mathfrak{sp}(2N+2)_k\oplus\mathfrak{so}(4N+4)_1}{\mathfrak{sp}(2N)_{k+1}}\oplus\mathfrak{u}(1) \; ,
\end{equation}
where $\mathfrak{so}(4N+4)_1$ encodes the $4N+4$ free fermions.\\

The generators of the $\mathfrak{sp}(2N)$ algebra are described using a double negative index notation, see \cite{Girardi:1981qz}, $J^{(a,b)}$ for $a,b=\pm 1,\ldots, \pm N$, and such that 
\begin{equation}\label{eqn:COND}
J^{(a,b)}=-\text{sgn}(ab)J^{(-b,-a)} \; .
\end{equation}
Indeed, a general $\mathfrak{sp}(2N)$ matrix $A$ is of the form
\begin{equation}\label{eqn:A}
A=
\left( \begin{array}{c|c}
~B~ & C \\  \hline  D & -B^t 
\end{array} \right) \; ,
\end{equation}
where $B$, $C$, $D$ are $N\times N$ matrices such that $C^t=C$ and $D^t=D$. This matches the notation above if we identify $J^{(a,b)}$ for $a,b>0$ as the generators of $B$, $J^{(a,b)}$ for $a>0$, $b<0$ as the generators of $C$, and $J^{(a,b)}$ for $a<0$, $b>0$ as the generators of $D$. \\

Note that $M\in\mathfrak{gl}(N)$ can be embedded into $\mathfrak{sp}(2N)$ by $M\oplus(-M^t)$. In particular $\mathfrak{u}(N)\subset\mathfrak{sp}(2N)$ by $U\oplus U^*$ for $U\in\mathfrak{u}(n)$. The $N^2$ generators of $\mathfrak{u}(N)$ correspond then to $J^{(a,b)}$ for $a,b>0$, and the vector representation $\mathbf{2N}$ of $\mathfrak{sp}(2N)$ splits into $\mathbf{N}\oplus\bar{\mathbf{N}}$, where $\mathbf{N}$, $\bar{\mathbf{N}}$ are the fundamental and anti-fundamental representations of $\mathfrak{u}(N)$, respectively. In terms of the adjoint generators, they correspond to $J^{(a,b)}$ for $a,b>0$ and $a,b<0$. Since this embedding is diagonal, it can be trivially extended to the group level.\\

The structure constants are given in \cite{Girardi:1981qz} as
\begin{equation}
f^{(ab)(cd)(ef)}=\text{sgn}(bc)\left(\delta^b_{~c}\delta^e_{~a}\delta^f_{~d}+\delta^d_{~a}\delta^e_{~-b}\delta^f_{~-c}+\delta^d_{~-b}\delta^e_{~a}\delta^f_{~-c}+\delta^c_{~-a}\delta^e_{~-b}\delta^f_{~d}\right) \; ,
\end{equation}
and the Cartan generators are $H_a=J^{(a,a)}$ for $a>0$. The adjoint representation of $\mathfrak{sp}(2N+2)$ branches as
\begin{equation}
\mathfrak{sp}(2N+2)=\mathfrak{sp}(2N)\oplus\mathfrak{sp}(2)\oplus(\mathbf{2N},\mathbf{2}) \; ,
\end{equation}
with $\mathfrak{sp}(2)\cong\mathfrak{sl}(2)$. The regular embedding $\mathfrak{sp}(2N)\longhookrightarrow\mathfrak{sp}(2N+2)$ is defined such that the generators $J^{(a,b)}$ split as
\begin{equation}
\begin{array}{ccc}
\mathfrak{sp}(2N): & J^{(i,j)} & \text{for }i,j=\pm 2,\ldots,\pm(N+1)\\
\mathfrak{sp}(2): & J^{(1,1)}, J^{(1,-1)}, J^{(-1,1)} &\\
(\mathbf{2N},\mathbf{2}): & J^{(i,\alpha)} & \text{for }i=\pm2,\ldots,\pm(N+1),~\alpha=\pm.
\end{array}
\end{equation}
Furthermore, by embedding $\mathfrak{sp}(2N)$ in $\mathfrak{so}(4N+4)$, the free fermions transform as
\begin{equation}\label{eqn:fermions}
2\times(\mathbf{2N}) \oplus 4\times (\mathbf{1}) \; ,
\end{equation}
with respect to $\mathfrak{sp}(2N)$.\\

%%%%%%%%%%%%%%%%%%%%%%%%%%%%%%%%%%%%%%%%%%%%%%%%%%%%%%%%%%%%%%%%%%%%%%
\subsection{The $\mathfrak{su}(2)_{\pm}$ subalgebras}\label{app:Rsymm}
%%%%%%%%%%%%%%%%%%%%%%%%%%%%%%%%%%%%%%%%%%%%%%%%%%%%%%%%%%%%%%%%%%%%%%

There are $4N+4$ free fermions in the set of generators of the coset, transforming as in \eqref{eqn:fermions}. These divide into three fermions $\psi^{(1,1)}$, $\psi^{(-1,1)}$, and $\psi^{(1,-1)}$, from $\mathfrak{sp}(2)^{(1)}$, one fermion from the $\mathfrak{u}(1)^{(1)}$, and 4$N$ fermions in the $(\mathbf{2N},\mathbf{2})$ of $\mathfrak{sp}(2N)\oplus\mathfrak{sp}(2)$ denoted $\psi^{i,\alpha}$ for $i=\pm2,...,\pm(N+1)$ and $\alpha=\pm$. The $4N$ fermions $\psi^{i,\alpha}$ satisfy the OPE
\begin{equation}
\psi^{i,\alpha}(z)\psi^{j,\beta}(w)\sim\frac{\delta_{ij}\delta_{\alpha\beta}}{z-w} \; ,
\end{equation}
and an $\mathfrak{su}(2)_N$ affine algebra can be constructed from these by defining the generators
\begin{equation}
\tilde{K}^{\alpha\beta}=\sum_{i,j}\Omega_{ij}(\psi^{i,\alpha}\psi^{j,\beta}) \; .
\end{equation}
In the double negative index notation, this current has the form
\begin{equation}
\tilde{K}^{\alpha\beta}=\sum_i \text{sgn}(i)(\psi^{i,\alpha}\psi^{-i,\beta}) = \sum_{i=1}^N\left( (\psi^{i,\alpha}\psi^{-i,\beta})- (\psi^{-i,\alpha}\psi^{i,\beta}) \right) \; ,
\end{equation}
with respect to which the $4N$ fermions transform in $2N\times{\bf 2}$. From this, the two $\mathfrak{su}(2)_{k^{\pm}}$ forming the R-symmetry of the $\mathcal{N}=4$ superconformal algebra can be constructed just as in \cite{Gaberdiel:2013vva}. Decoupling the fermions $\psi^{1,1}$, $\psi^{-1,1}$, and $\psi^{1,-1}$ from $\mathfrak{sp}(2)_{k+N+2}$ we get a bosonic $\mathfrak{sp}(2)_{k+N}$, since $\check{h}_{\mathfrak{su}(2)}=2$, whose generators we denote $J$. Then construct $\tilde{J}=J-\tilde{K}$, which generate $\mathfrak{su}(2)_{k}$, and with respect to which the $4N$ fermions transform trivially. We have therefore constructed the subalgebra
\begin{equation}
 \mathfrak{su}(2)_k\oplus\mathfrak{su}(2)_N~~\text{generated by }\tilde{J}\oplus\tilde{K} \; .
\end{equation}
Gathering the three free fermions from the decoupling above, together with the fermion from $\mathfrak{u}(1)^{(1)}$, we form $\mathfrak{so}(4)_1\cong\mathfrak{su}(2)_1\oplus\mathfrak{su}(2)_1$, with respect to which the free fermions by construction transform in the $(\mathbf{2},\mathbf{2})$. Add each one of the $\mathfrak{su}(2)$ factors to $\tilde{J}$ and $\tilde{K}$, to obtain
\begin{equation}
 \mathfrak{su}(2)_{N+1}\oplus\mathfrak{su}(2)_{k+1} \; .
\end{equation}

%%%%%%%%%%%%%%%%%%%%%%%%%%%%%%%%%%%%%%%%%%%%%%%%%%%%%%%%%%%%%%%%%
\subsection{Selection rules}\label{BPS}
%%%%%%%%%%%%%%%%%%%%%%%%%%%%%%%%%%%%%%%%%%%%%%%%%%%%%%%%%%%%%%%%%

For highest weights $\Lambda^+$, $\Lambda^-$ of $\mathfrak{sp}(2N+2)$ and $\mathfrak{sp}(2N)$ respectivelly, decomposing as $(\Lambda^+_1,\ldots,\Lambda^+_{N+1})$ and $(\Lambda^-_1,\ldots,\Lambda^-_N)$ in a basis of fundamental weights $\omega_i=\epsilon_1+\ldots+\epsilon_i$, where $\left\{\epsilon_i\right\}_{i=1}^{N+1}$ is an orthonormal basis of the weight space, the selection rules state that $\mathcal{P}\Lambda^+-\Lambda^-\in\mathcal{P}\mathcal{Q}^{N+1}$, where $\mathcal{P}\mathcal{Q}^{N+1}$ is the projection of the root lattice of $\mathfrak{sp}(2N+2)$. Given the embedding above, the highest root $\theta$ projects to zero, the first root $\alpha_1$ is the $\mathfrak{sp}(2)$ root, while the other simple roots $\alpha_i$ for $i=2,\ldots,N+1$ span the denominator $\mathfrak{sp}(2N)$ root system. The projection of $\theta$ to zero then allows to express $\alpha_1$ in terms of the other simple roots, since
\begin{equation}
\theta=2\sum_{i=1}^N\alpha_i+\alpha_{N+1}\xrightarrow{\mathcal{P}} 0 \; ,
\end{equation}
that is, $\alpha_1=-\sum_{i=2}^N\alpha_i-\frac{1}{2}\alpha_{N+1}$ upon projection. In the simple root basis $\Lambda^+$ takes the expression
\begin{align}\label{ROOTBASIS}
\nonumber \Lambda^+ & =\left(\Lambda^+_1+\ldots\Lambda^+_{N+1}\right)\alpha_1+\left(\Lambda^+_1+2\Lambda^+_2+\ldots 2\Lambda^+_{N+1}\right)\alpha_2+\\
& +\ldots +\left(\Lambda^+_1+2\Lambda^+_2+3\Lambda^+_3+\ldots+(N+1)\Lambda^+_{N+1}\right)\frac{1}{2}\alpha_{N+1} \; ,
\end{align}
which upon projection yields
\begin{align}
\nonumber \mathcal{P}\Lambda^+ & =\left(\Lambda^+_2+\ldots+\Lambda^+_{N+1}\right)\alpha_2+\left(\Lambda^+_2+2\Lambda^+_3+\ldots+3\Lambda_{N+1}^+\right)\alpha_3\\
& +\ldots +\left(\Lambda_2^++2\Lambda_3^+\ldots+N\Lambda_{N+1}^+\right)\frac{1}{2}\alpha_{N+1}\; .
\end{align}
It is clear that $\alpha_1$ disappeared from the expression and this may now be compared with $\Lambda^-$. The root lattice $\mathcal{Q}$ projects as follows: a general element in $\mathcal{Q}^{N+1}$ decomposing as $\alpha=\sum_{i=1}^{N+1}n_i\alpha_i$, with $n_i\in\mathbb{Z}$, is mapped by the projection above into
\begin{align}
\nonumber \mathcal{P}\alpha & =n_1\left(-\sum_{i=2}^N\alpha_i-\frac{1}{2}\alpha_{N+1}\right)+\sum_{i=2}^{N+1}n_i\alpha_i\\
& = \sum_{i=2}^{N}\left(n_i-n_1\right)\alpha_i+\left(2n_{N+1}-n_1\right)\frac{1}{2}\alpha_{N+1} \; .
\end{align}
A general element of $\mathcal{P}\mathcal{Q}^{N+1}$ is then of the form
 \begin{equation}
\alpha=\sum_{i=2}^{N}n_i\alpha_i+n_{N+1}\frac{1}{2}\alpha_{N+1}=\sum_{i=2}^{N+1}n_i\check{\alpha}_i \; ,
\end{equation}
for $n_i\in\mathbb{Z}$. We made use of the fact that the co-roots have the form $\check{\alpha_i}=\alpha_i$ for $i=1,\ldots,N$, and $\check{\alpha}_{N+1}=\frac{1}{2}\alpha_{N+1}$. We have therefore established that 
\begin{equation}
\mathcal{P}\mathcal{Q}^{N+1}\cong\check{\mathcal{Q}}^N\; ,
\end{equation}
where $\check{\mathcal{Q}^N}$ is the co-root lattice of $\mathfrak{sp}(2N)$. It is now clear that the selection rules are always trivially satisfied: the weight $\mathcal{P}\Lambda^+$ is easily seen from \eqref{ROOTBASIS} to be an element of $\check{\mathcal{Q}}^N$, and the same happens with $\Lambda^-$. This ultimately stems from the fact that the weight lattice and the co-root lattice of $\mathfrak{sp}(2N+2)$ are isomorphic. Then $\mathcal{P}\Lambda^+-\Lambda^-$ also lies in $\check{\mathcal{Q}}^N$, and therefore in $\mathcal{P}\mathcal{Q}^{N+1}$. The selection rules are therefore trivial.\\

%%%%%%%%%%%%%%%%%%%%%%%%%%%%%%%%%%%%%%%
\subsection{Field identifications}
%%%%%%%%%%%%%%%%%%%%%%%%%%%%%%%%%%%%%%%

The group of outer automorphisms of $\mathfrak{sp}(2N)$ is $\mathcal{O}=\mathbb{Z}_2=\left\{\mathds{1},J\right\}$, where $J$ may be defined by its action on an affine weight:
\begin{equation}
J\cdot\left[\Lambda_0; \Lambda_1,\ldots,\Lambda_N\right]=\left[\Lambda_N; \Lambda_{N-1},\ldots,\Lambda_1,\Lambda_0\right] \; .
\end{equation}
Following the usual rules to determine the branching of the outer automorphisms, and the inner product $(\omega_i,\omega_j)=\frac{1}{2}\text{min}(i,j)$, we compute
\begin{align}
\left(J\cdot\omega_0,\Lambda\right) & =\frac{1}{2}|\Lambda | \\
\left(\mathds{1}\cdot\omega_0,\Lambda\right) & =0 \; ,
\end{align}
as well as 
\begin{align}
\left(\tilde{J}\cdot\hat{\omega}_0,\mathcal{P}\Lambda\right) & =\frac{1}{2}|\Lambda |-\frac{1}{2}\Lambda_1\\
\left(\tilde{\mathds{1}}\cdot\hat{\omega}_0,\mathcal{P}\Lambda\right) & =0 \; ,
\end{align}
where $J$, $\tilde{J}$ are the non-trivial outer automorphisms of $\mathfrak{sp}(2N+2)$ and $\mathfrak{sp}(2N)$, respectively, and $|\Lambda |=\sum_{i=1}^{N+1}i\Lambda_i$. Given the branching condition 
\begin{equation}
\left(A\cdot\omega_0,\Lambda\right)-\left(\tilde{A}\cdot\omega_0,\mathcal{P}\Lambda\right)=0 \text{ mod 1},
\end{equation}
for $A=\mathds{1},J$ and $\tilde{A}=\tilde{\mathds{1}},\tilde{J}$, it is easy to see that the condition is only satisfied for $\mathds{1}\mapsto\tilde{\mathds{1}}$. Therefore there are no non-trivial branchings, and the field identifications are trivial.

%%%%%%%%%%%%%%%%%%%%%%%%%%%%%
\subsection{BPS states}\label{app:BPS}
%%%%%%%%%%%%%%%%%%%%%%%%%%%%%

In terms of partition coefficients $l_1\geq l_2\geq\ldots\geq l_N\geq 0$, or Dynkin labels $\Lambda_i\in\mathbb{N}^0$, which are related by
\begin{equation}\label{eqn:partCoeff}
l_i = \sum_{j=i}^N \Lambda_j \; ,
\end{equation}
or equivalently
\begin{equation}
\Lambda_i = l_i-l_{i+1} \; ,
\end{equation}
the quadratic Casimir of $\mathfrak{sp}(2N)$ is given by
\begin{align}
\begin{aligned}
C^{(N)}(\Lambda) & =  \frac{1}{4}\sum_{i=1}^N\Lambda_i\left(\sum_{j=i+1}^Ni\Lambda_j+\sum_{j=1}^i j\Lambda_j+i(2N+1-i)\right) \\
& = \frac{1}{4}\left(\sum_{i=1}^Nl_i^2+2\sum_{i=1}^N(N-1+i)l_i\right) \;.
\end{aligned}
\end{align}

For integrable highest weights $\Lambda^+$ and $\Lambda^-$ of $\mathfrak{sp}(2N+2)$ and $\mathfrak{sp}(2N)$, respectivelly, the conformal dimension of the coset representation $(\Lambda^+;\Lambda^-)$ is given by
\begin{equation}\label{H}
h(\Lambda^+;\Lambda^-)=\frac{C^{(N+1)}(\Lambda^+)-C^{(N)}(\Lambda^-)}{k+N+2}+n \; ,
\end{equation}
where $n$ is an integer specifying at which level $\Lambda^+$ appears in the decomposition of $\Lambda^-$.\\

The BPS bound for representations of the $\mathcal{N}=4$ superconformal algebra is \cite{Gaberdiel:2013vva}
\begin{equation}
h(l^{\pm},u)\geq\frac{1}{k^++k^-}\left[k^+l^-+k^-l^++u^2+(l^+-l^-)^2\right],
\end{equation}
for two representations $l^{\pm}$ of the R-symmetry algebras $\mathfrak{su}(2)_{\pm}$, and the $\mathfrak{u}(1)$ representation $u$ (which we will take as $u=0$, as implicitly assumed already in \eqref{H}). The coset representations $\left(\langle i,0,\ldots,0\rangle;0\right)$ have conformal dimension 
\begin{equation}
h(\langle i,0,\ldots,0\rangle;0)=\frac{i}{2(k+N+2)}\left(\frac{i}{2}+N+1\right) \; ,
\end{equation}
and therefore saturate the BPS bound for $l^+=i/2$ and $l^-=0$. On the other hand, the representations $(0;v_j)\equiv\left(0;\langle 0,\ldots,1,\ldots,0\rangle\right)$, for which the only non-zero Dynkin label is $\Lambda^-_j=1$, have conformal dimension
\begin{equation}
h(0;\langle 0,\ldots,1,\ldots,0\rangle)=\frac{j}{2(k+N+2)}\left(\frac{j}{2}+k+1\right) \; ,
\end{equation}
and saturate the BPS bound for $l^+=0$, $l^-=j/2$.\\

Note that, in the 't Hooft limit, and for $v\equiv\langle 1,0,\ldots,0\rangle$, we have
\begin{alignat}{3}
h(v;0) & = \frac{N+3/2}{2(k+N+2)} & \simeq \frac{N}{2(k+N)} & \simeq \frac{\lambda}{2} \\
h(0;v) & = \frac{k+3/2}{2(k+N+2)} & \simeq \frac{k}{2(k+N)} & \simeq \frac{1-\lambda}{2} \; .
\end{alignat}
In particular, for $\lambda =0$,
\begin{equation}
h(0;v)\simeq \frac{1}{2} \; .
\end{equation}
The other BPS representations $(0;v_j)$ appear in the $j$-th tensor power of $(0;v)$ and have conformal dimension $j/2$ for $\lambda =0$.

%%%%%%%%%%%%%%%%%%%%%%%%%%%%%%%%%%%%%%%%%%%%%%%%%%%%%%%%%
\subsection{Wedge characters}\label{app:NULL}
%%%%%%%%%%%%%%%%%%%%%%%%%%%%%%%%%%%%%%%%%%%%%%%%%%%%%%%%%

In the following, we present circumstantial evidence in favour of the claim that
\begin{equation}
\chi_{(0;\Xi)}^{\text{(wedge})} = \chi_{(0;\Lambda)}^{(\text{wedge})} \; ,
\end{equation}
where $\chi_{(0;\Xi)}^{\text{(wedge})}$ is the wedge character of the representation $(0;\Xi)$ of the Sp($2N$) vector model, and $\chi_{(0;\Lambda)}^{(\text{wedge})}$ is the wedge character of the representation $(0;\Lambda)$ of the U($N$) vector model. No rigorous proof is provided in general. The wedge character of $(0;v)$ is
\begin{equation}
\chi^{\text{(wedge})}_{(0;v)}(q,y)=\frac{\left(y^{1/2}+y^{-1/2}\right)q^{1/2}+2q}{1-q}\; .
\end{equation}
Note that this character is equal to the character $\chi^{\text{(wedge)}}_{(0;f)}$ of the U($N$) model, c.f.\ \cite{Gaberdiel:2013vva}. As in previously considered models \cite{Candu:2012jq,Gaberdiel:2011nt,Gaberdiel:2011zw}, we claim that the decoupling of null states at large $N$ is taken care of by requiring that the (wedge) fusion rules of the representations of the Sp(2$N$) vector model are simply given by the fusion rules of the U($N$) vector model. This then ensures that the vacuum representation does not appear in the decomposition of $(0;v)\otimes_{\text{f}} (0;v)$. This is analoguous to the requirement in \cite{Candu:2012jq,Gaberdiel:2011nt,Gaberdiel:2011zw} that the vacuum representation does not appear in the decomposition of $(0;f)\otimes_{\text{f}}(0;\bar{f})$. The explicit fusion rules are then
\begin{equation}
(0;\Xi^1)\otimes_{\text{f}}(0;\Xi^2)=\bigoplus_{\substack{\Xi\\ \vert\Xi\vert=\vert\Xi^1\vert+\vert\Xi^2\vert}} (0;\Xi) \; ,
\end{equation}
where $\vert \Xi\vert = \sum_i i\; \Xi_i$ is the number of boxes of the corresponding Young diagram. In conclusion, since $\chi^{\text{(wedge)}}_{(0;v)}=\chi^{\text{(wedge)}}_{(0;f)}$ and all the fusion rules coincide, all the wedge characters must also be the same,
\begin{equation}
\chi^{\text{(wedge)}}_{(0;\Xi)}=\chi^{\text{(wedge)}}_{(0;\Lambda)} \; ,
\end{equation}
for any Sp($2N$) representation $\Xi$, where $\Lambda$ is the U($N$) representation with Dynkin labels $\Lambda_i=\Xi_i$.\\

%From the point of view of AdS, the representation $(0;v)$ corresponds to the degrees of freedom of two massless real scalars and two massive Dirac fermions. The CFT fusion rules given above correspond \\

%%%%%%%%%%%%%%%%%%%%%%%%%%%%%%%%%%%%%%%%%%%%%%%%%%%%%%%%%%%%
\section{The large level limit of the coset}\label{app:kINFTY}
%%%%%%%%%%%%%%%%%%%%%%%%%%%%%%%%%%%%%%%%%%%%%%%%%%%%%%%%%%%%

%%%%%%%%%%%%%%%%%%%%%%%%%%%%%%%%%
\subsection{Untwisted sector}
%%%%%%%%%%%%%%%%%%%%%%%%%%%%%%%%%

 The untwisted sector of the vector model is captured by the $k\rightarrow\infty$ limit of the $(0;\Lambda)$ closed subsector of the coset representations. The coset character of these representations is denoted $b_{(0;\Lambda)}^{N,k}(q)$ and can be obtained from
\begin{equation}
 \text{ch}_0^{N+1,k}(\iota (v),q)\cdot\theta(v,q)=\sum_{\Lambda}b_{(0;\Lambda)}^{N,k}(q)\text{ch}_{\Lambda}^{N,k+1}(v,q) \; ,
\end{equation}
 where $\iota (v)$ is the embedding of the $\mathfrak{sp}(2N)$ weights into $\mathfrak{sp}(2N+2)$, and $\theta (v,q)$ is the character of the $4N+4$ free fermions. Also, $\text{ch}_0^{N+1,k}$ is the character of the trivial representation of $\mathfrak{sp}(2N+2)_k$, and $\text{ch}_{\Lambda}^{N,k+1}$ is the character of the $\Lambda$ representation of $\mathfrak{sp}(2N)_{k+1}$. Using the Kac-Weyl formula (see \cite{DiFrancesco1997}), and given the Sp$(2N)$ roots $\pm\epsilon_i\pm\epsilon_j$ for $i\neq j$, and $\pm2\epsilon_i$, $i,j=1,\ldots,N$, the expression above becomes in the $k\rightarrow\infty$ limit:
\begin{align}\label{eqn:CHAR}
 \nonumber\sum_{\Lambda}a_{(0;\Lambda)}(q)ch_{\Lambda}^N(v,q) & =\frac{\theta (v,q)} {\prod_{n=1}^{\infty} \prod_{i=2}^{N+1}(1-v_iv_1q^n)(1-v_i^{-1}v_1q^n)(1-v_i^{-1}v_1^{-1}q^n)}\\
&\times \frac{1}{(1-v_iv_1^{-1}q^n)(1-v_1^2q^n)(1-v_1^{-2}q^n)(1-q^n)},
\end{align}
 where $a_{(0;\Lambda)}\cong b_{(0;\Lambda)}^{N,k}$ in the $k\rightarrow\infty$ limit. On the other hand, given the embedding specified in Appendix \ref{app:coset},
\begin{equation}
 \theta(v,q)=\prod_{n=1}^{\infty}\prod_{i=2}^{N+1}(1+v_iq^{n+1/2})^2(1+v_i^{-1}q^{n+1/2})^2(1+q^{n+1/2})^4 \; .
\end{equation}
 Not keeping track of the $\mathfrak{sp}(2)$ eigenvalues by setting $v_1=1$, we finally find
\begin{align}\label{COSET}
 \nonumber\sum_{\Lambda}a_{(0;\Lambda)}(q)ch_{\Lambda}^N(v,q) & =\prod_{n=1}^{\infty}\prod_{i=2}^{N+1}\frac{(1+v_iq^{n+1/2})^2(1+v_i^{-1}q^{n+1/2})^2(1+q^{n+1/2})^4} {(1-v_iq^n)^2(1-v_i^{-1}q^n)^2(1-q^n)^4} \; ,
\end{align}
 which leads to the identification of the $k\rightarrow\infty$ limit of the $(0;\Lambda)$ subsector of the coset theory as the $\mathfrak{sp}(2N)$ continuous orbifold of $4N+4$ free bosons and fermions transforming in the $2\times(\mathbf{2N})\oplus 4\times(\mathbf{1})$ representation.

%%%%%%%%%%%%%%%%%%%%%%%%%%%%%%%%%%%%%%
\subsection{Twisted sectors}
%%%%%%%%%%%%%%%%%%%%%%%%%%%%%%%%%%%%%%

 The Cartan torus of $\mathfrak{sp}(2N)$ may be chosen as $\text{diag}(z_1,\ldots,z_N,z_1^{-1},\ldots,z_N^{-1})$, for $z_j=e^{2\pi i\alpha_j}$ and $\alpha_j\in\left[-\frac{1}{2},\frac{1}{2}\right]$. In order to label conjugacy classes, we divide out the action of the Weyl group $W=S_N\ltimes\mathbb{Z}_2^N$, which consists in exchanging the twists $\alpha_j$ and reversing their sign. Then conjugacy classes are then labeled by $\alpha=\left[\alpha_1,\ldots,\alpha_N\right]$ satisfying
\begin{equation}
\frac{1}{2}\geq\alpha_1\geq\ldots\geq\alpha_N\geq 0 \; .
\end{equation}
 The conformal dimension of the $\alpha$-twisted sector is given by $h(\alpha)=\sum_{i=1}^N|\alpha_i|=\sum_{i=1}^{N}\alpha_i$. For $m=0,\ldots,N$, the twisted sector ground states correspond to the coset representations $\left(\Lambda^{(m)}_+;\Lambda^{(m)}_-\right)$ where
\begin{align}\label{eqn:twistedReps}
 \Lambda^{(m)}_+ & = \langle\Lambda_1,\ldots,\Lambda_m,0,\ldots,0\rangle \nonumber \\
 \Lambda^{(m)}_- & = \langle\Lambda_1,\ldots,\Lambda_m,0,\ldots,0\rangle \; .
\end{align}
The corresponding twist is claimed to be
\begin{equation}
 \alpha=\frac{1}{k+N+2}\left[\sum_{i=1}^{m}\Lambda_i,\sum_{i=2}^{m}\Lambda_i,\ldots,\Lambda_m,0,\ldots,0\right] \; .
\end{equation}
 in the $k\rightarrow\infty$ limit. Note that, by definition of the partition coefficients $l_i$, $i=1,\ldots N$, given in \eqref{eqn:partCoeff}, we have
\begin{equation}\label{eqn:twist}
\alpha_i = \frac{l_i}{k+N+2} \; .
\end{equation}
 In the following subsections we give some evidence supporting this identification.

%%%%%%%%%%%%%%%%%%%%%%%%%%%%%%%%%%%%%%%%%
\subsubsection{Conformal dimensions}
%%%%%%%%%%%%%%%%%%%%%%%%%%%%%%%%%%%%%%%%%

 The corresponding conformal dimensions can be seen to match: the conformal dimension of the coset representation is given by (in terms of partition coefficients)
\begin{align}
 \nonumber h\left(\Lambda^{(m)}_+;\Lambda^{(m)}_-\right)  = \frac{1}{k+N+2} & \left(C^{(N+1)}(\Lambda_+^{(m)})-C^{(N)}(\Lambda_-^{(m)})\right)  \\
 \nonumber = \frac{1}{2(k+N+2)} & \left(\sum_{i=1}^{N+1}l_i^2+2\sum_{i=1}^{N+1}(N-i+2)l_i\right.  \\
 &  -\left.\sum_{i=1}^Nl_i^2-2\sum_{i=1}^{N}(N-i+1)l_i\right) \; .
\end{align}
 Using the explicit form of \eqref{eqn:twistedReps}, together with \eqref{eqn:twist}, it becomes
\begin{align}
 h\left(\Lambda^{(m)}_+;\Lambda^{(m)}_-\right) &=\frac{1}{2(k+N+2)}\left(4\sum_{i=1}^m l_i-2\sum_{i=1}^m l_i\right)\\
& = \sum_{i=1}^m\alpha_i,
\end{align}
 which coincides with the twisted sector conformal dimension since $\alpha_i=0$ for $i>m$.

%%%%%%%%%%%%%%%%%%%%%%%%%%%%%%%%%%%%%%%%%%%%%%%%%%
\subsubsection{Fermionic excitation spectrum}
%%%%%%%%%%%%%%%%%%%%%%%%%%%%%%%%%%%%%%%%%%%%%%%%%%

 The fusion of a coset representation $(\Lambda_+;\Lambda_-)$ with the minimal representation $(0;v)$ has the following form:
\begin{equation}
(\Lambda_+;\Lambda_-)\otimes(0;v)=(\Lambda_+;\Lambda_-\otimes v) \; ,
\end{equation}
where $\Lambda_-\otimes v$ decomposes as
\begin{equation}
 \Lambda_-\otimes v=\bigoplus_{\epsilon=\pm}\bigoplus_{r=1}^N\Lambda^{(r,\epsilon)} \; ,
\end{equation}
 with
\begin{equation}
\Lambda_j^{(r,\epsilon)}=
\left\{
\begin{array}{ccl}
\Lambda_j+\epsilon && j = r-1\\
\Lambda_j-\epsilon && j =  r\\
\Lambda_j && \text{otherwise} \; .
\end{array}
\right.
\end{equation}
 With respect to the original coset state, the partition coefficients change as
\begin{equation}
l_r\rightarrow l_r-\epsilon \; ,
\end{equation}
 while all the others remain the same. Then the conformal dimension of the fusion product differs from the original one by
\begin{align}
 \nonumber \delta h^{(r)} &= h(\Lambda_+;\Lambda_-)-h(\Lambda_+;\Lambda^{(r,\epsilon)}) \\
 \nonumber &= \frac{1}{2}+\frac{1}{k+N+2}\left(C^{(N)}(\Lambda_-)-C^{(N)}(\Lambda_-^{(r,\epsilon)})\right)\\
 \nonumber &= \frac{1}{2}+\frac{1}{k+N+2}\left(-\epsilon l_r+\frac{1}{2}-\epsilon(N-r+1)\right)\\
&\cong \frac{1}{2}-\epsilon \alpha_r,
\end{align}
 where in the last line we have used \eqref{eqn:twist}, and took the $k\rightarrow\infty$ limit. It is then clear that each of the channels of the fusion of a $(\Lambda^{(m)}_+;\Lambda_-^{(m)})$ state with $(0;v)$ corresponds to a state twisted by $\pm\alpha_r$, and therefore this state is indeed the twisted sector ground state for $\alpha=[\alpha_1,\ldots,\alpha_N]$.

\section{Higher spin algebras}\label{app:HS}
%%%%%%%%%%%%%%%%%%%%%%%%%%%%%%%%%%%%%%%%%%%%%%%%%%%%%%%%%%%%%

Higher spin theories in AdS$_3$ are described as Chern-Simons theories with a higher spin algebra as gauge algebra. In this section we mainly follow \cite{Candu:2013fta}, \cite{Candu:2014yva}. The super Lie algebra $\mathfrak{shs}_2[\mu]$ is the tensor product of two different components: a gravitational part, and an internal part. The gravitational part consists of the associative algebra defined as
\begin{equation}
sB[\mu]=\frac{U(\mathfrak{osp}(1,2))}{\langle C_2-\frac{1}{4}\mu(\mu-1)\mathds{1}\rangle}\cong sB[1-\mu] \; ,
\end{equation}
where $C_2$ is the second Casimir of $\mathfrak{osp}(1,2)$. This algebra can be faithfully realised in terms of an associative algebra spanned by the oscillators $\hat{y}_{\alpha}$, $\alpha=1,2$ and an operator $k$, together with the identity element $\mathds{1}$, satisfying
\begin{alignat}{3}\label{eqn:RELATIONS}
\left[ \hat{y}_{\alpha}, \hat{y}_{\beta} \right] = 2i\epsilon_{\alpha\beta}(\mathds{1}+ \nu k), & \qquad & k\hat{y}_{\alpha}=-\hat{y}_{\alpha}k, & \qquad & k^2=\mathds{1} \; ,
\end{alignat}
with $\nu = 2\mu -1$, so that the generators of the AdS$_3$ superalgebra $\mathfrak{osp}(1,2)$ are
\begin{alignat}{3}
T_{\alpha\beta}=\frac{1}{4i}\left\lbrace \hat{y}_{\alpha},\hat{y}_{\beta}\right\rbrace, & \qquad & \hat{y}_{\alpha} \; ,
\end{alignat}
and the second Casimir is
\begin{equation}
C_2=-\frac{1}{2}T_{\alpha\beta}T^{\alpha\beta}-\frac{i}{4}\hat{y}_{\alpha}\hat{y}^{\alpha} \; ,
\end{equation}
which can be seen to equal $\frac{1}{4}\mu(\mu-1)$ automatically, using the oscillator realisation. Indices are raised and lowered using $\epsilon_{\alpha\beta}$.\\

Using the grading $\vert\hat{y}_{\alpha}\vert = 1$, $\vert k\vert =\vert \mathds{1}\vert = 0$, this associative algebra can be turned into a super Lie algebra $\mathfrak{shs}[\mu]$ by defining a bracket as
\begin{equation}
\left[ a,b\right]_{\pm} = ab - (-1)^{\vert a\vert \vert b\vert}ba \; ,
\end{equation}
with $a,b\in sB[\mu]$, and by quotienting out the central element of $sB[\mu]$. The \emph{spin} of a given element of $\mathfrak{shs}[\mu]$ is defined as its eigenvalue under the adjoint action of the Cartan generator of the gravitational $\mathfrak{sl}(2)$ subalgebra generated by $T_{\alpha\beta}$. We can immediately see that $\hat{y}_1$ has spin 1/2, whereas $\hat{y}_2$ has spin $-1/2$, since they form a doublet. Higher powers of the oscillators in the associative algebra are associated with higher spin fields, transforming in a certain representation of the $\mathfrak{sl}(2)$ subalgebra.\\

An extended associative algebra $sB_2[\mu]$ can be obtained by tensoring the gravitational $sB[\mu]$ with a matrix algebra describing Chan-Paton degrees of freedom:
\begin{equation}
sB_M[\mu] = sB[\mu]\otimes \text{Mat}(2,\mathbb{C})\; ,
\end{equation}
where $\text{Mat}(2,\mathbb{C})$ is the usual algebra of complex $2\times 2$ matrices. This part of the extended algebra does not change the properties of the elements with respect to the gravitational part, and only adds some degeneracy. Proceeding as before, we obtain the Lie superalgebra $\mathfrak{shs}_2[\mu]$, given by
\begin{equation}
\mathfrak{shs}_2[\mu] = \mathds{1}\otimes\mathfrak{sl}(2) \oplus \mathfrak{shs}[\mu]\otimes\mathds{1}_2 \oplus \mathfrak{shs}[\mu]\otimes\mathfrak{sl}(2) \; .
\end{equation}

An important result obtained in \cite{Gaberdiel:2013vva} is that $\mathfrak{shs}_2[\mu]$ contains the $\mathcal{N}=4$ super Lie algebra $D(2,1;\alpha)$ as a subalgebra, with
\begin{equation}
\alpha = \frac{\mu}{\mu-1} = \frac{\nu + 1}{\nu-1}\; .
\end{equation} 
The basis elements of $D(2,1;\alpha)$ are realised in $\mathfrak{shs}_2[\mu]$ as
\begin{align}\label{eqn:D210}
\begin{aligned}
L_0 & = \frac{1}{8i}\left(\hat{y}_1\hat{y}_2+\hat{y}_2\hat{y}_1\right)\otimes\mathbf{1}_2 \\
L_{ 1} & = \frac{1}{4i}\hat{y}_1\hat{y}_1\otimes\mathds{1}_2 \\
L_{-1} & = \frac{1}{4i}\hat{y}_2\hat{y}_2\otimes\mathds{1}_2 \\
A^{\pm, i}_0 & = \frac{1}{2}\left( 1\pm k\right)\otimes \sigma^i \\
G_{r}^{++} & = e^{\pi i/4} \hat{y}_{\alpha_r}k\otimes E_{12}  \\
G_{r}^{--} & = - e^{\pi i/4} \hat{y}_{\alpha_r} k\otimes E_{21}  \\
G_{r}^{-+} & = - \frac{e^{\pi i/4}}{2} \left[ \hat{y}_{\alpha_r}\otimes \mathbf{1}_2 + \hat{y}_{\alpha_r}k\otimes(E_{11}-E_{22}) \right]  \\
G_{r}^{+-} & =  \frac{e^{\pi i/4}}{2} \left[ \hat{y}_{\alpha_r}\otimes \mathbf{1}_2 - \hat{y}_{\alpha_r}k\otimes\left(E_{11}-E_{22}\right) \right]   \; ,
\end{aligned}
\end{align}
where $E_{ab}$ is the matrix whose only non-zero entry (equal to 1) is in the $a,b$ position, $\sigma^i$ are the Pauli matrices, and $\alpha_r=3/2-r$, $r=\pm 1/2$.\\

%%%%%%%%%%%%%%%%%%%%%%%%%%%%%%%%%%%%%%%%%%%%%%%%%%%%%
\subsection{Truncations of higher spin algebras}
%%%%%%%%%%%%%%%%%%%%%%%%%%%%%%%%%%%%%%%%%%%%%%%%%%%%%

A graded automorphism $\tau$ of a super Lie algebra $L$ is defined as a linear invertible map of $L$ onto itself satisfying
\begin{align}\label{eqn:AUTO}
\tau(\left[a_1, a_2\right]_{\pm}) & = \left[\tau(a_1),\tau(a_2)\right]_{\pm} \\
\vert \tau(a_1)\vert & =\vert a_1\vert \; ,
\end{align}
for $a_1,a_2\in L$. All the elements $a\in L$ satisfying
\begin{equation}
\tau(a)=a \; ,
\end{equation}
form a subalgebra, by virtue of \eqref{eqn:AUTO}. The automorphism is called involutive if $\tau^2=1$. An anti-automorphism of second class $\eta$ (henceforth shortened to \emph{anti-automorphism}) is a linear invertible map of a graded associative algebra $A$ onto itself, which satisfies
\begin{equation}\label{eqn:ANTIAUTO}
\eta(a_1\cdot a_2) = (-1)^{\vert a_1\vert \vert a_2\vert}\eta(a_2)\cdot\eta(a_1) \; .
\end{equation}
By endowing $A$ with a bracket and turning it into a super Lie algebra $L_A$, and if $\eta$ preserves the grading $\vert\eta(a)\vert=\vert a\vert$, then
\begin{equation}
\eta\left( \left[a_1,a_2\right]_{\pm}\right)=-\left[\eta(a_1),\eta(a_2)\right]_{\pm} \; ,
\end{equation}
and therefore an automorphism $\tau$ of $L_A$ can be constructed as
\begin{equation}
\tau = -\eta \; .
\end{equation}

Consistent higher spin theories can be obtained from the theory with gauge algebra $\mathfrak{shs}_2[\mu]$ by the use of automorphisms $\tau$ of $\mathfrak{shs}_2[\mu]$, or anti-automorphisms $\eta$ of the associative algebra $sB_2[\mu]$, which preserve the gravitational $\mathfrak{sl}(2)$, and also that satisfy the consistency condition $\tau(k)=k$, $\eta(k)=k$. The automorphisms of the higher spin algebra define the real forms of the higher spin algebra, whereas anti-automorphisms of the associative algebra give rise to consistent theories with a truncated spectrum of massless fields, see \cite{Vasiliev:1999ba,Vasiliev:1986qx,Konstein:1989ij,Prokushkin:1998bq}.

%%%%%%%%%%%%%%%%%%%%%%%%%%%%%%%%%%%%%%%%%%%%%%%%%%%%%%%%%%%%
\section{Matching one-loop partition functions}\label{app:DUAL}
%%%%%%%%%%%%%%%%%%%%%%%%%%%%%%%%%%%%%%%%%%%%%%%%%%%%%%%%%%%%

In this section we compute and match the thermal partition function of AdS$_3$ Chern-Simons theory with symmetry algebra $\mathfrak{shs}^{\text{sp}}_2[\mu]$, supplemented with two real scalars, with the partition function of the dual coset theory in the 't Hooft limit. 

%%%%%%%%%%%%%%%%%%%%%%%%%%%%%%%%%%%%%%%%%%%
\subsection{Coset partition functions}
%%%%%%%%%%%%%%%%%%%%%%%%%%%%%%%%%%%%%%%%%%%
The character of the coset representation $(\Lambda_+;\Lambda_-)$ is defined as
\begin{equation}
b^{N,k}_{(\Lambda_+;\Lambda_-)}(q,y_{\pm})=\text{tr}_{(\Lambda_+;\Lambda_-)}\left(q^{L_0}\text{exp}^{\left[J_0^I\text{tr}(t^IH_+)+K_0^I\text{tr}(t^IH_-)\right]}\right) \; ,
\end{equation}
where $H_{\pm}$ are arbitrary elements of a Cartan subalgebra of the horizontal subalgebra of the affine $\mathfrak{sp}(2)_+\oplus\mathfrak{sp}(2)_-$ generated by the spin-1 currents $J^I$, $K^I$, with eigenvalues $z^i_{\pm}$, $i=1,\ldots,M$, in the fundamental representation. From these characters we construct the partition function
\begin{equation}
 \mathcal{Z}(q,y_{\pm})=\vert q^{-\frac{c}{24}}\vert^2\sum_{\left[(\Lambda^+;\Lambda^-)\right]}\vert b^{N,k}_{(\Lambda_+;\Lambda_-)}(q,y_{\pm})\vert^2 \; .
\end{equation}
We are interested in the 't Hooft limit for which $N,k\rightarrow\infty$ with
\begin{equation}
 \lambda=\frac{k^-}{k^++k^-}\simeq\frac{N}{k+N} \; ,
\end{equation}
where $k^{\pm}$ are the levels of the affine algebras $\mathfrak{sp}(2)_{\pm}$. The coset characters are obtained from the character decomposition
\begin{equation}\label{eqn:BRANCH}
 \text{ch}_{\Lambda_+}^{2N+2,k}(q,\iota_1(y_+,v))\theta(q,\iota_2(y_{\pm},v))=\sum_{\Lambda_-}b^{N,k}_{(\Lambda_+;\Lambda_-)}(q,y_{\pm})\text{ch}_{\Lambda_-}^{2N,k+1}(q,v) \; ,
\end{equation}
where $\text{ch}_{\Lambda_+}^{2N+2,k}$ is the character of the $\Lambda_+$ representation of $\mathfrak{sp}(2N+2)_k$, whereas $\iota_{1,2}$ denote the embeddings of the numerator into the denominator algebras, with $v$ a $\mathfrak{sp}(2N)$ matrix with eigenvalues $v^a$, and $y_+$ a $\mathfrak{sp}(2)$ matrix with eigenvalues $y_+^i$.\\

In the 't Hooft limit the branching identity \eqref{eqn:BRANCH} can be recast into the form
\begin{align}
\begin{aligned}
 \sum_{\Lambda_-}a_{(\Lambda_+;\Lambda_-)}(q,y_{\pm}) & \; \text{ch}_{\Lambda_-}^{2N}(q,v) =  \text{ch}_{\Lambda_+}^{2N+2}(q,\iota_1(y_+,v)) \\
 & \times\prod_{a=1}^{2N}\prod_{i=\pm\frac{1}{2}}\prod_{n=1}^{\infty}\frac{(1+y_-^{-i}v^aq^{n-1/2})(1+y_-^iv^{-a}q^{n-1/2})}{(1-y_+^{-i}v^aq^n)(1-y_+^iv^{-a}q^n)} \; ,
 \end{aligned}
\end{align}
where
\begin{equation}
 b^{N,k}_{(\Lambda_+;\Lambda_-)}(q,y_{\pm})\cong q^{\frac{1}{2\kappa}\left[\text{Cas}^{2N+2}(\Lambda_+)-\text{Cas}^{2N(}\Lambda_-)\right]}a_{(\Lambda_+;\Lambda_-)}(q,y_{\pm}) \; ,
\end{equation}
\\
in the 't Hooft limit, and all the characters are now just regular Weyl characters.\\

For $\Lambda_+=0$, this expresses $a_{(0;\Lambda_-)}$ as the multiplicity of the $\Lambda_-$ representation of $\mathfrak{sp}(2N)$ in a system of free bosons and fermions transforming as given. In particular, for $\Lambda_-=0$, this corresponds to the vacuum character $a_{(0;0)}(q,y_{\pm})$, encoding the chiral algebra of the cosets. For a general $\Lambda_-$ this multiplicity may be found using the methods of \cite{Candu:2012jq}, \cite{Candu:2013fta}. The right hand side of the expression above is the character of an infinite dimensional vector space spanned by vectors of the form
\begin{equation}
 \prod_{k=1}^{n_{\psi}}\psi_{-r_k-1/2}^{a_ki_k}\prod_{l=1}^{n_j}j_{-s_l-1}^{b_lj_l}\vert 0\rangle \; ,
\end{equation}
for $r_k,s_l\in\mathbb{N}^0$. Since the action of $\mathfrak{gl}(\infty\vert\infty)_+$ on these mode numbers and the action of $\mathfrak{sp}(2N)$ on the indices commute, the multiplicity of $\Lambda_-$ will naturally be a character of $\mathfrak{gl}(\infty\vert\infty)_+$. For a fixed number of fields $n_{\psi}$, $n_j$, a given $\mathfrak{sp}(2N)$ representation $\Lambda_-$ such that $\vert\Lambda_-\vert=n_{\psi}+n_j$ will appear with multiplicity 0 or 1, where $\vert\Lambda_-\vert$ is the number of boxes of the corresponding Young diagram. This multiplicity will be 1 only if there is a Young supertableau of shape $\Lambda_-$ with even entries from $\left\{2s_l+2\right\}$, and odd entries from $\left\{2r_k+1\right\}$. Summing over all possible mode numbers, and multiplying by all possible invariant states, the total contribution of these to the subspace transforming in $\
\Lambda_-$ is
\begin{equation}
 a_{(0;\Lambda_-)}(q,y_{\pm})= a_{(0;0)}(q,y_{\pm})\text{sch}_{\Lambda^t_-}(\mathcal{U}_1) \; ,
\end{equation}
where 
\begin{equation}
\text{sch}_{\Lambda}(\mathcal{U}(h))=\sum_{T\in STab_{\Lambda}}\prod_{i\in T}q^{h+\frac{i}{2}} \; ,
\end{equation}
is the supercharacter in the $\Lambda_-$ representation of the diagonal matrix $\mathcal{U}(h)\in$ GL$(\infty\vert\infty)_+$ with entries
\begin{equation}
 \mathcal{U}(h)_{jj}=(-1)^jq^{h+\frac{j}{2}} \; ,
\end{equation}
and $\mathcal{U}_1=\mathcal{U}(h=1/2)$. \\

In line with previous cases (see \cite{Candu:2012jq}, but also \cite{Gaberdiel:2011zw,Gaberdiel:2011nt,Candu:2013fta}), we claim that the emergence of null vectors in the 't Hooft limit is taken care of by requiring that the fusion rules of coset representations become effectively the U($N$) tensor rules once fundamental and antifundamental representations are decoupled (see \cite{Gaberdiel:2011nt} for a similar situation). This is implemented by demanding that the total number of boxes does not decrease after fusion, that is
\begin{equation}\label{eqn:FUSION}
 \Lambda_+\otimes_{\text{f}}\Lambda_-=\bigoplus_{\substack{\Lambda\\\vert\Lambda\vert=\vert\Lambda_+\vert+\vert\Lambda_-\vert}}\Lambda \; ,
\end{equation}
with $\vert\Lambda\vert$ denoting the total number of boxes in the respective Young diagram. If we denote $c_{\Phi\Psi}^{\Lambda}$ the Clebsch-Gordan coefficients for $\mathfrak{sp}$, such that
\begin{equation}
 \text{ch}_{\Lambda}^{2N+2}(\iota_1(y_+,v))=\sum_{\Phi,\Psi}c_{\Phi\Psi}^{\Lambda}\text{ch}_{\Phi}^{2N}(v)\text{ch}_{\Psi}^2(y_+) \; ,
\end{equation}
then using the fusion relation \eqref{eqn:FUSION}, which implies the factorisation of the Clebsch-Gordan coefficients, we obtain the following expression for a general coset character in the 't Hooft limit,
\begin{equation}
 a_{(\Lambda_+;\Lambda_-)}(q,y_{\pm})=\sum_{\Phi,\Psi,\Pi}c_{\Pi\Phi}^{\Lambda_+}c_{\Psi\Phi}^{\Lambda_-}\text{ch}^2_{\Pi}(y_+)a_{(0;\Psi)}(q,y_{\pm}) \; .
\end{equation}
Rewriting the expression above in terms of the vacuum coset character, we get
\begin{equation}
 a_{(\Lambda_+;\Lambda_-)}(q,y_{\pm})=\sum_{\Phi,\Psi,\Pi}c_{\Pi\Phi}^{\Lambda_+}c_{\Psi\Phi}^{\Lambda_-}\text{ch}^2_{\Pi}(y_+)\text{sch}_{\Psi^t}(\mathcal{U}_1)a_{(0;0)}(q,y_{\pm}) \; ,
\end{equation}
on which we can use the properties of the Clebsch-Gordan coefficients, and the identity (see \cite{Candu:2013fta})
\begin{equation}
 \text{sch}_{\Lambda}(\mathcal{U}_0)=\sum_{\Phi,\Pi}c_{\Pi\Phi}^{\Lambda}\text{ch}^2_{\Pi}(z_+)\text{sch}_{\Phi^t}(\mathcal{U}_1) \; ,
\end{equation}
where $\mathcal{U}_0=\mathcal{U}(h=0)$, to obtain the simplified expression
\begin{equation}
 a_{(\Lambda_+;\Lambda_-)}(q,y_{\pm})=\text{sch}_{\Lambda_+}(\mathcal{U}_0)\text{sch}_{\Lambda_-^t}(\mathcal{U}_1)a_{(0;0)}(q,y_{\pm})\; .
\end{equation}

Finally, in the 't Hooft limit the overall multiplying factor simplifies to
\begin{equation}
  q^{\frac{1}{2\kappa}\left[\text{Cas}^{N+2}(\Lambda_+)-\text{Cas}^N(\Lambda_-)\right]}\cong q^{\frac{\lambda}{2}(\vert\Lambda_+\vert-\vert\Lambda_-\vert)} \; ,
\end{equation}
which we can absorb in the entries of the matrices $\mathcal{U}_0$, $\mathcal{U}_1$ by defining
\begin{align}
 \text{sch}_{\Lambda_+}(\mathcal{U}_+) & = q^{\frac{\lambda}{2}\vert\Lambda_+\vert}\text{sch}_{\Lambda_+}(\mathcal{U}_0) \\  
\text{sch}_{\Lambda_-}(\mathcal{U}_-) & = q^{-\frac{\lambda}{2}\vert\Lambda_-\vert}\text{sch}_{\Lambda_-}(\mathcal{U}_1) \; .
\end{align}
We are now able to write the partition function of the coset in the 't Hooft limit:
\begin{equation}\label{eqn:PART}
 \mathcal{Z}^{\text{'t Hooft}}(q,y_{\pm})=\sum_{\Lambda_+,\Lambda_-}\vert\text{sch}_{\Lambda_+}(\mathcal{U}_+)\text{sch}_{\Lambda_-}(\mathcal{U}_-)a_{(0;0)}(q,y_{\pm})\vert^2 \; .
\end{equation}

%%%%%%%%%%%%%%%%%%%%%%%%%%%%%%%%%%%%%%%%%%%%
\subsection{AdS$_3$ partition function}
%%%%%%%%%%%%%%%%%%%%%%%%%%%%%%%%%%%%%%%%%%%%

As proven before, the chiral spectrum of the proposed coset dual, counted by $a_{(0;0)}(q,y_{\pm})$, coincides with the gauge sector of the AdS$_3$ truncated extended higher spin theory:
\begin{equation}
 \mathcal{Z}_{\text{gauge}}(q,y_{\pm})=\vert a_{(0;0)}(q,y_{\pm})\vert^2 \; .
\end{equation}
Following \cite{Candu:2013fta}, full correspondence is achieved by adding matter fields, as in \ref{subsec:MASSIVE}. The total matter contribution to the partition function in thermal AdS$_3$ is then
\begin{equation}
\mathcal{Z}_{\text{matter}}(q,y_{\pm})=\mathcal{Z}^+_{\text{matter}}(q,y_{\pm})\mathcal{Z}^-_{\text{matter}}(q,y_{\pm}) \; ,
\end{equation}
with
\begin{equation}
\mathcal{Z}^{\pm}_{\text{matter}}(q,y_{\pm})=q^{h_{\pm}}\bar{q}^{h_{\pm}}\prod_{i,j=\pm\frac{1}{2}}\prod_{m,n=0}^{\infty}\frac{(1+y_{\mp}^i\bar{y}_{\pm}^jq^{m+1/2}\bar{q}^{n})(1+y_{\pm}^i\bar{y}_{\mp}^jq^{m}\bar{q}^{n+1/2})}{(1-y_{\pm}^i\bar{y}_{\pm}^jq^{m}\bar{q}^{n})(1-y_{\mp}^i\bar{y}_{\mp}^jq^{m+1/2}\bar{q}^{n+1/2})} \; ,
\end{equation}
where
\begin{alignat}{3}
  h_+=\frac{\mu}{2}, \; & \qquad & h_-=\frac{1-\mu}{2} \; .
 \end{alignat}
We can now use the GL$(\infty\vert\infty)$ supermatrix
\begin{align}
\begin{aligned}
 \mathcal{U}_{\pm}=q^{h_{\pm}}\text{diag}(y_{\pm}^{1/2},y_{\pm}^{-1/2}, & -y_{\mp}^{1/2}q^{1/2},-y_{\mp}^{-1/2}q^{1/2}, \\
& y_{\pm}^{1/2}q^{1},y_{\pm}^{-1/2}q^{1},-y_{\mp}^{1/2}q^{3/2},-y_{\mp}^{-1/2}q^{3/2},\ldots) \; ,
\end{aligned}
\end{align}
together with the Cauchy identity, to write the above expression as
\begin{equation}
 \mathcal{Z}^{\pm}_{\text{matter}}(q,y_{\pm})=\sum_{\Lambda}\text{sch}_{\Lambda}(\mathcal{U}_{\pm})\overline{\text{sch}_{\Lambda}(\mathcal{U}_{\pm})} \; .
\end{equation}
It is now evident we can match the higher spin partition function
\begin{equation}
\mathcal{Z}^{\text{1-loop}}_{\text{AdS}_3}(q,y_{\pm})=\mathcal{Z}_{\text{gauge}}(q,y_{\pm})\mathcal{Z}_{\text{matter}}^+(q,y_{\pm})\mathcal{Z}_{\text{matter}}^-(q,y_{\pm}) \; ,
\end{equation}
with the 't Hooft limit of the coset partition function \eqref{eqn:PART}, if we identify $\lambda=\mu$.

%%%%%%%%%%%%%%%%%%%%%%%%%%%%%%%%%%%%
\bibliographystyle{JHEP}
%\bibliography{evenN4}
%%%%%%%%%%%%%%%%%%%%%%%%%%%%%%%%%%%%
\providecommand{\href}[2]{#2}\begingroup\raggedright\endgroup

\end{document}